\documentclass[12pt]{article}

\usepackage{amsmath}
\usepackage{amsfonts}
\usepackage{graphicx}
\usepackage{caption}
\usepackage{subcaption}

\usepackage{authblk}
\usepackage{amssymb}

\usepackage{epstopdf}
\usepackage{epsfig}
\usepackage{hyperref}

\usepackage{color}

\date{}

\begin{document}

\title{Radial perturbations of the scalarized EGB black holes}

\author[1]{Jose Luis Bl\'azquez-Salcedo \thanks{\href{mailto:jose.blazquez.salcedo@uni-oldenburg.de}{jose.blazquez.salcedo@uni-oldenburg.de}}}

\author[2,3]{Daniela D. Doneva
\thanks{\href{mailto:daniela.doneva@uni-tuebingen.de }{daniela.doneva@uni-tuebingen.de }}}

\author[1]{Jutta Kunz \thanks{\href{mailto:jutta.kunz@uni-oldenburg.de}{jutta.kunz@uni-oldenburg.de}}} 

\author[2,4,5]{Stoytcho S. Yazadjiev \thanks{\href{mailto:yazad@phys.uni-sofia.bg}{yazad@phys.uni-sofia.bg}}}

\affil[1]{Institut f\"ur  Physik, Universit\"at Oldenburg, Postfach 2503,
D-26111 Oldenburg, Germany}
\affil[2]{Theoretical Astrophysics, Eberhard Karls University of T\"ubingen, T\"ubingen 72076, Germany}
\affil[3]{INRNE - Bulgarian Academy of Sciences, 1784  Sofia, Bulgaria}
\affil[4]{Department of Theoretical Physics, Faculty of Physics, Sofia University, Sofia 1164, Bulgaria}
\affil[5]{Institute of Mathematics and Informatics, Bulgarian Academy of Sciences, Acad. G. Bonchev Street 8, Sofia 1113, Bulgaria}

\maketitle

\begin{abstract}
Recently a new class of scalarized black holes in Einstein-Gauss-Bonnet (EGB) theories was discovered. What is special for these black hole solutions is that the scalarization is not due to the presence of matter, but {it is induced} by the curvature of spacetime itself. Moreover, more than one branch of scalarized solutions can bifurcate from the Schwarzschild branch, and these scalarized branches are characterized by the number of nodes of the scalar field. The next step is to consider the linear stability of these solutions, which is particularly important due to the fact that the Schwarzschild black holes lose stability at the first point of bifurcation. Therefore we here study in detail the radial perturbations of the scalarized EGB black holes. The results show that all branches with a nontrivial scalar field with one or more nodes are unstable. The stability of the solutions on the fundamental branch, whose scalar field has no radial nodes, depends on the particular choice of the coupling function between the scalar field and the Gauss-Bonnet invariant. We consider two particular cases based on the previous studies of the background solutions. If this coupling has the form used in \cite{Doneva:2017bvd} the fundamental branch of solutions is stable, except for very small masses. In the case of a coupling function quadratic in the scalar field \cite{Silva:2017uqg}, though, the whole fundamental branch is unstable.
\end{abstract}
\section{Introduction}
Very recently new black holes with a nontrivial scalar field have been constructed in the extended scalar-tensor-Gauss-Bonnet (ESTGB) theories \cite{Doneva:2017bvd,Silva:2017uqg,Antoniou:2017acq,Antoniou:2017hxj}. What is interesting in these results is the presence of non-uniqueness of the solutions -- in addition to the pure general relativistic solution, that exists in the whole domain of the parameter space, for a certain range of parameters new branches of solutions with a nontrivial scalar field are present. These branches can be characterized by the number of nodes of the scalar field. In fact, besides the fundamental branch which possesses no nodes of the scalar field, there arises a whole sequence of radially excited branches \footnote{Note, that radially excited black hole solutions were known before only in Einstein-Maxwell-Chern-Simons theory \cite{Blazquez-Salcedo:2015kja} {and in scalar tensor theories in the presence of a nonlinear electromagnetic field \cite{Stefanov:2007eq,Doneva:2010ke}}.}. Moreover, the Schwarzschild solution loses stability at the point, where the first nontrivial branch bifurcates from it. Then the fundamental branch of scalarized black holes could represent the stable one. This would represent a direct analogy with the spontaneous scalarization of neutron stars in the standard scalar-tensor theories considered in \cite{Damour:1993hw}, and also with the scalarized black holes in scalar-tensor theories in the presence of nonlinear matter sources \cite{Stefanov:2007eq,Doneva:2010ke,Cardoso:2013fwa,Cardoso:2013opa}. The main difference with respect to those results, though, is that in the ESTGB case the scalar field is not sourced by matter, but instead by the curvature of spacetime through the Gauss-Bonnet scalar. In fact such spontaneous scalarization in ESTGB theories is observed also for neutron stars \cite{Doneva:2017duq,Silva:2017uqg}.

The ESTGB theories are very interesting on their own because of the following reasons. Their theoretical motivation comes from attempts to quantize gravity and the fact that pure general relativity is not a renormalizable theory. A way to cure this problem is to supplement the Einstein-Hilbert action  with all possible algebraic curvature invariants of second order \cite{Stelle:1976gc}. A serious problem that appears, though, is that the resulting field equations are of order higher than two, which leads to Ostrogradski instability and to the appearance of ghosts. However, this can be avoided in the special case when the additional dynamical scalar field is non-minimally coupled to a special combination of the second order invariants, namely the Gauss-Bonnet invariant, since the resulting field equations in this case are of second order \cite{Berti:2015itd}. These are exactly the  ESTGB theories. 

One of the most studied class of ESTGB theories are the dilatonic EGB theories, where the coupling function between the scalar field and the Gauss-Bonnet invariant has the form $\alpha e^{2\gamma \varphi}$, where $\alpha$ and $\gamma$ are arbitrary constants, and the scalar field potential is set to zero. Black holes in dilatonic EGB theories have been extensively studied, both in the perturbative and non-perturbative regime and also including rapid rotation \cite{Mignemi:1992nt}--\cite{Kleihaus:2015aje}. In contrast with the dilatonic EGB theories, the considered class of ESTGB theories in \cite{Doneva:2017bvd,Silva:2017uqg,Antoniou:2017acq,Antoniou:2017hxj} is characterized by a coupling function that can lead to non-uniqueness of the solutions and scalarization/descalarization.

Stability of black holes in dilatonic EGB theories was examined in \cite{Kanti:1997br,Torii:1998gm,Pani:2009wy,Ayzenberg:2013wua,Blazquez-Salcedo:2016enn,Blazquez-Salcedo:2017txk} and it was shown that the primary branch of black holes is stable, while the secondary branch, that appears for sufficiently strong dilaton coupling, is unstable. The linear stability of the scalarized black holes obtained in \cite{Doneva:2017bvd,Silva:2017uqg,Antoniou:2017acq,Antoniou:2017hxj} has not been studied yet. It was already proved, though, in \cite{Doneva:2017bvd,Silva:2017uqg,Myung:2018iyq} that the Schwarzschild solution loses stability at the point of bifurcation, where the  first nontrivial branch of solutions appears. Examining the stability of the nontrivial branches of solutions is much more involved and represents the goal of the present paper. Based on thermodynamical analysis, it has been argued in \cite{Doneva:2017bvd} that, for the particular coupling function considered there, the fundamental ESTGB black hole branch should be the stable one,
whereas all radially excited branches should be unstable. Of course in order to check this hypothesis more rigorously, one has to examine the linear stability of the branches of scalarized black holes, as done below.

In Section II the field equations used to obtain the background solutions are presented. The radial perturbations are examined in Section III, while the lengthy formulae are given in a separate Appendix. The  results for the stability of the scalarized black holes are presented in Section IV. The paper ends with Conclusions.

\section{Field equations}
The action of ESTGB theories in vacuum, in its general form, can be written as
\begin{eqnarray}
S=&&\frac{1}{16\pi}\int d^4x \sqrt{-g} 
\Big[R - 2\nabla_\mu \varphi \nabla^\mu \varphi - V(\varphi) 
+ \lambda^2 f(\varphi){\cal R}^2_{GB} \Big] ,\label{eq:quadratic}
\end{eqnarray}
where $R$ is the Ricci scalar with respect to the spacetime metric $g_{\mu\nu}$, $\varphi$ is the scalar field, $V(\varphi)$ and  $f(\varphi)$ are the potential and the coupling function that  depend on $\varphi$ only, $\lambda$ is the Gauss-Bonnet coupling constant that has  dimension of $length$. The  Gauss-Bonnet invariant ${\cal R}^2_{GB}$ is defined as ${\cal R}^2_{GB}=R^2 - 4 R_{\mu\nu} R^{\mu\nu} + R_{\mu\nu\alpha\beta}R^{\mu\nu\alpha\beta}$ where $R$ is the Ricci scalar, $R_{\mu\nu}$ is the Ricci tensor and $R_{\mu\nu\alpha\beta}$ is the Riemann tensor. 

We will consider static and spherically symmetric spacetimes as well as static and spherically symmetric scalar field configurations. In addition, we will concentrate on the case of zero scalar field potential $V(\varphi)=0$. Thus, the spacetime metric can be written in the following form
\begin{eqnarray}
ds^2= - e^{2\Phi(r)}dt^2 + e^{2\Lambda(r)} dr^2 + r^2 (d\theta^2 + \sin^2\theta d\phi^2 ). 
\end{eqnarray}   
After varying the action and performing a dimensional reduction of the resulting field equations one can obtain the following system of ordinary differential equations (more details can be found in \cite{Doneva:2017bvd})
\begin{eqnarray}
&&\frac{2}{r}\left[1 +  \frac{2}{r} (1-3e^{-2\Lambda})  \Psi_{r}  \right]  \frac{d\Lambda}{dr} + \frac{(e^{2\Lambda}-1)}{r^2} 
- \frac{4}{r^2}(1-e^{-2\Lambda}) \frac{d\Psi_{r}}{dr} - \left( \frac{d\varphi}{dr}\right)^2=0, \label{DRFE1}\\ && \nonumber \\
&&\frac{2}{r}\left[1 +  \frac{2}{r} (1-3e^{-2\Lambda})  \Psi_{r}  \right]  \frac{d\Phi}{dr} - \frac{(e^{2\Lambda}-1)}{r^2} - \left( \frac{d\varphi}{dr}\right)^2=0,\label{DRFE2}\\ && \nonumber \\
&& \frac{d^2\Phi}{dr^2} + \left(\frac{d\Phi}{dr} + \frac{1}{r}\right)\left(\frac{d\Phi}{dr} - \frac{d\Lambda}{dr}\right)  + \frac{4e^{-2\Lambda}}{r}\left[3\frac{d\Phi}{dr}\frac{d\Lambda}{dr} - \frac{d^2\Phi}{dr^2} - \left(\frac{d\Phi}{dr}\right)^2 \right]\Psi_{r} 
\nonumber \\ 
&& \hspace{0.9cm} - \frac{4e^{-2\Lambda}}{r}\frac{d\Phi}{dr} \frac{d\Psi_r}{dr} + \left(\frac{d\varphi}{dr}\right)^2=0, \label{DRFE3}\\ && \nonumber \\
&& \frac{d^2\varphi}{dr^2}  + \left(\frac{d\Phi}{dr} \nonumber - \frac{d\Lambda}{dr} + \frac{2}{r}\right)\frac{d\varphi}{dr} \nonumber \\ 
&& \hspace{0.9cm} - \frac{2\lambda^2}{r^2} \frac{df(\varphi)}{d\varphi}\Big\{(1-e^{-2\Lambda})\left[\frac{d^2\Phi}{dr^2} + \frac{d\Phi}{dr} \left(\frac{d\Phi}{dr} - \frac{d\Lambda}{dr}\right)\right]    + 2e^{-2\Lambda}\frac{d\Phi}{dr} \frac{d\Lambda}{dr}\Big\} =0, \label{DRFE4}
\end{eqnarray}
where
\begin{eqnarray}
\Psi_{r}=\lambda^2 \frac{df(\varphi)}{d\varphi} \frac{d\varphi}{dr}.
\end{eqnarray}

Furthermore, we assume zero cosmological value of the scalar field $\varphi|_{r\rightarrow\infty}\equiv\varphi_{\infty}=0$, and the coupling function  $f(\varphi)$ is chosen such that it satisfies the conditions $\frac{df}{d\varphi}(0)=0$ and $b^2=\frac{d^2f}{d\varphi^2}(0)>0$.  Without loss of generality we can  set $b=1$, which can be achieved after rescaling of the coupling parameter $\lambda\to b\lambda$ and redefinition of the function $f\to b^{-2}f$. Another observation one can make is that the field equations do not depend on $f(\varphi)$ itself, but only on its derivative with respect to $\varphi$ which leaves the freedom to choose $f(0)=0$. 

After performing an expansion of the reduced field equations at the horizon and requiring regularity of the metric functions and the scalar field, one finds \cite{Doneva:2017bvd} that black hole solutions 
with a real scalar field
exist only when the following condition is fulfilled
\begin{equation}
r_H^4 > 24 \lambda^4 \left(\frac{df}{d\varphi}(\varphi_{H})\right)^2, \label{eq:BC_sqrt_rh}
\end{equation}
where $r_H$ is the radius of the black hole horizon and $\varphi_{H}$ is the value of the scalar field at the horizon.

The boundary conditions are derived by the requirement for asymptotic flatness 
\begin{eqnarray}
\Phi|_{r\rightarrow\infty} \rightarrow 0, \;\;  \Lambda|_{r\rightarrow\infty} \rightarrow 0,\;\; \varphi|_{r\rightarrow\infty} \rightarrow 0\;\;.   \label{eq:BH_inf}
\end{eqnarray} 
On the other hand, the very existence of a black hole horizon requires 
\begin{eqnarray}
e^{2\Phi}|_{r\rightarrow r_H} \rightarrow 0, \;\; e^{-2\Lambda}|_{r\rightarrow r_H} \rightarrow 0. \label{eq:BC_rh}
\end{eqnarray}

In the present paper we will concentrate mainly on the following coupling function
\begin{equation}
f(\varphi)=  \frac{1}{12} \left(1- e^{-6\varphi^2}\right), \label{eq:coupling_function}
\end{equation}
since it was shown in \cite{Doneva:2017bvd} that it can produce non-negligible deviations from  pure GR. In addition, eq. \eqref{eq:coupling_function} is quite similar to the coupling function employed in the studies of spontaneous scalarization of neutron stars \cite{Damour:1993hw}. 

In the last part of the paper we will also present results for the case of the quadratic potential previously considered in \cite{Silva:2017uqg}
 \begin{equation}
f(\varphi)=  \frac{1}{2} \varphi^2. \label{eq:coupling_function_quad}
\end{equation}
It is worth noting that in the case of small scalar field, the coupling \eqref{eq:coupling_function_quad} is the leading term of the coupling \eqref{eq:coupling_function}, $f(\varphi)= \frac{1}{12} \left(1- e^{-6\varphi^2}\right) \approx \frac{1}{2} \varphi^2 + O(\varphi^4)$, and both couplings will share many features in the small $\varphi$ domain.

\section{Radial Perturbations}

\subsection{Ansatz and equations}

{We consider time dependent radial perturbations over the spherically symmetric and static background black holes obtained after solving the reduced system of equations \eqref{DRFE1}--\eqref{DRFE4}}
\begin{eqnarray} 
\label{metric_pert}
ds^2 &=& -e^{2\Phi(r)+\epsilon F_t(r,t)}dt^2 + e^{2\Lambda(r) + \epsilon F_r(r,t)} dr^2 + r^2 (d\theta^2 + \sin^2\theta d\phi^2), \nonumber \\
\varphi &=& \varphi_0(r) + \epsilon \varphi_1(r,t),
\end{eqnarray}
with $\epsilon$ {being the control parameter of the perturbations}. The field equations result in three second order differential equations and two algebraic constraints on the first derivatives \cite{Torii:1998gm}. {However, the system can be simplified into a single second order differential equation }
\begin{equation}
\label{wave_eq}
g^2(r)  \frac{\partial^2\varphi_1}{\partial t^2}   - \frac{\partial^2\varphi_1}{\partial r^2} + C_1(r) \frac{\partial\varphi_1}{\partial r}  
+  U(r)  \varphi_1=0,
\end{equation}
{where the functions $U(r)$, $g(r)$ and $C_1(r)$ depend only  on the background metric and scalar field. Their expressions are given in the appendix \ref{appendix1}.}

In order to study the mode stability of the background configuration, we decompose the perturbation function $\varphi_1$ as
\begin{equation}
 \varphi_1(r,t) = \varphi_1(r) e^{i\omega t}
\end{equation}
{and we obtain the master equation for the eigenvalue problem, namely}
\begin{eqnarray}
\label{master_eq} 
\frac{d^2\varphi_1}{dr^2} = C_1(r) \frac{d\varphi_1}{dr} + \left[ U(r) - \omega^2 g^2(r) \right] \varphi_1(r).
\end{eqnarray} 

{The master equation (\ref{master_eq}) can be cast into the standard Schr\"odinger form by defining the function $Z(r)$:}
\begin{eqnarray}
\varphi_1(r) = C_0(r) Z(r), 
\end{eqnarray}
{where $C_0(r)$ is the solution of the following differential equation}
\begin{equation}
\frac{1}{C_0}\frac{dC_0}{dr}= C_1 - \frac{1}{g}\frac{dg}{dr}.
\end{equation}
{Thus we obtain}
\begin{eqnarray}
\label{master_eq_simp} 
\frac{d^2Z}{dR^2} = \left[ V(R) - \omega^2 \right] Z,
\end{eqnarray} 
where we have defined the tortoise coordinate $R$ and the effective potential as
\begin{eqnarray}
\frac{dR}{dr} &=& g, \label{eq:g}\\
V(R) &=&  \frac{1}{g^2} \left( U + \frac{C_1}{C_0}\frac{dC_0}{dr}-\frac{1}{C_0}\frac{d^2C_0}{dr^2} \right).
\label{eq:potential}
\end{eqnarray}

Since we are interested in the stability analysis of the background solutions, we will focus on perturbations with purely imaginary eigenfrequencies: $\omega = i\omega_I$. 

\subsection{Boundary conditions and numerical method}

We want the perturbation to be outgoing at infinity and ingoing at the horizon:
\begin{eqnarray}
Z \xrightarrow[r \to \infty]{} e^{i\omega(t-R)} = e^{-\omega_I (t-R)}, \\
Z \xrightarrow[r \to r_H]{} e^{i\omega(t+R)} = e^{-\omega_I (t+R)}.
\end{eqnarray}
These boundary conditions simplify a lot for unstable modes possessing $\omega_I < 0$, and it is straightforward to show that $Z|_{r=\infty}=Z|_{r=r_H}=0$ {(see e.g. \cite{Doneva:2010ke}).}

In order to obtain the unstable modes, we implement the following numerical procedure. The first step is to generate a background solution for a set of fixed values for $r_H$, $\lambda$ and the number of nodes of the scalar field. We make use of Colsys \cite{Ascher:1979iha} in order to integrate the equations. Typically the background solutions have a relative precision of $10^{-10}$ with around 1000 points on a grid in the compactified coordinate $x=1-r_H/r$. With these solutions we calculate numerically the coefficients of eq.~(\ref{master_eq}), which is the equation we solve, since it is slightly simpler than eq.~(\ref{master_eq_simp}). 

Once the coefficients are calculated, we follow a scheme similar to the one described in \cite{Motahar:2017blm} to obtain the bifurcation points. We define the quantity $\omega^2\equiv E$, and promote it to an auxiliary function $E(r)$. This function satisfies a trivial differential equation, $\frac{dE}{dr}=0$, which is added to eq.~(\ref{master_eq}) to form a system of differential equations. The three boundary conditions that we impose on this system are $\varphi_1|_{r=r_H}=0$, $\varphi_1|_{r=r_0}=1$ and $\varphi_1|_{r=\infty}=0$, where $r_H < r_0 < \infty$.  The procedure is automatized, allowing us to rapidly calculate the eigenfrequency for several thousands of solutions in the parameter space.

\section{Stability analysis}

Using the procedure described above we calculate the potential and the unstable modes of 
the Schwarzschild branch as well as the first few branches of scalarized black holes for the two coupling functions (\ref{eq:coupling_function})
and (\ref{eq:coupling_function_quad}).

\subsection{Exponential coupling function}
In the following we will focus on the exponential coupling function (\ref{eq:coupling_function}), employed in
\cite{Doneva:2017bvd}, discussing first the nonperturbative background solutions and subsequently the unstable modes.

\subsubsection{Background solutions}
\begin{figure}[]
	\centering
	\includegraphics[width=0.38\textwidth,angle=-90]{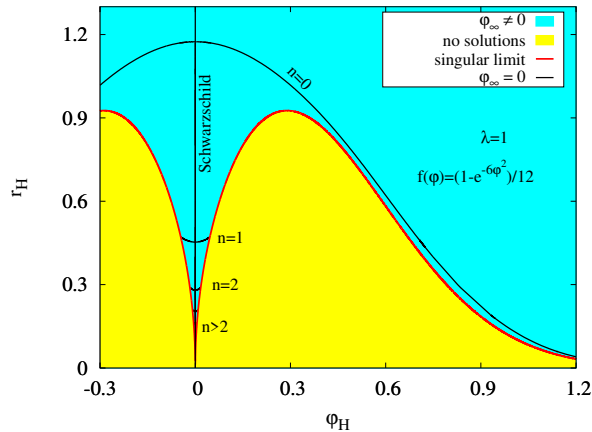}
	\caption{Domain of existence of black holes parametrized by $r_H$ and $\varphi_H$ for $\lambda=1$.}
	\label{Fig:domain}
\end{figure}

The domain of existence of these black holes is summarized in Fig.~\ref{Fig:domain}, where we show the space of solutions in the ($r_H$, $\varphi_H$) plane for $\lambda=1$ {\footnote{All dimensional quantities can be scaled with $\lambda$ in such a way that in the dimensionless representation of the field equations and their perturbations the parameter $\lambda$ does not appear explicitly. This is equivalent to consider $\lambda=1$ in our calculations. The exact dimensional values of the quantities will of course depend on the particular choice of $\lambda$.}}. 
The yellow area represents the region where condition \eqref{eq:BC_sqrt_rh} is not fulfilled, with the red line representing the saturation of this inequality (i.e., the singular limit).
 The area in cyan is filled by solutions that in general do not satisfy the condition $\varphi_{\infty}=0$. 
This condition is only satisfied for the solutions shown with black curves. Thus the $\varphi_{\infty}=0$ solutions form a system of branches bifurcating from the Schwarzschild solution (shown by the vertical line in Fig.~\ref{Fig:domain}). In the following we will only consider these branches of $\varphi_{\infty}=0$ solutions. 

Each branch of scalarized black holes can be characterized by the number of nodes of the scalar field as it extends along the radial coordinate. The fundamental branch possesses solutions without nodes ($n=0$), the first excited branch has solutions with one node ($n=1$), etc. Here we present results for the first six branches ($n=0...5$), although branches with higher number of nodes exist, presenting similar features.

\begin{figure}[]
	\centering
	\includegraphics[width=0.38\textwidth,angle=-90]{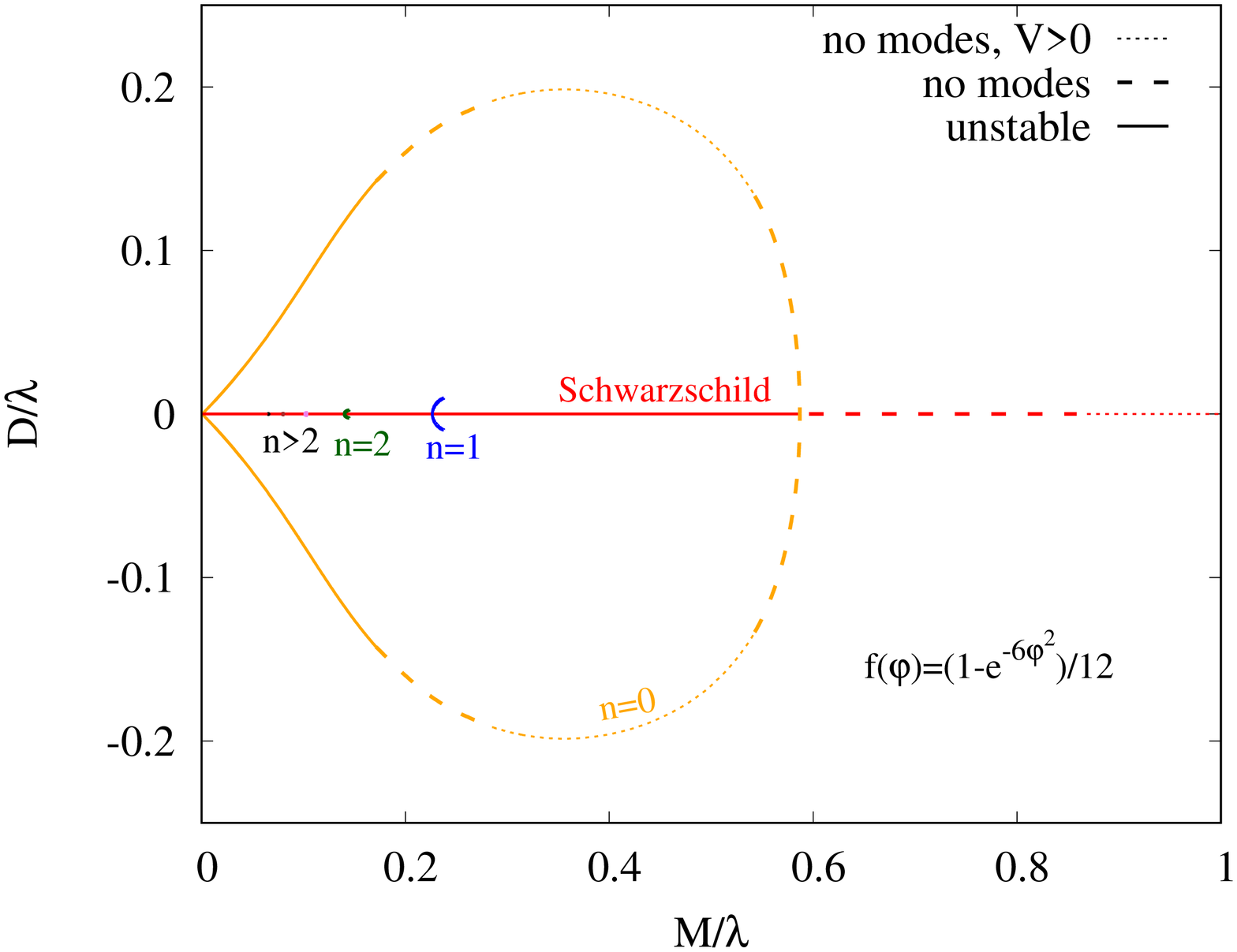}
    \includegraphics[width=0.38\textwidth,angle=-90]{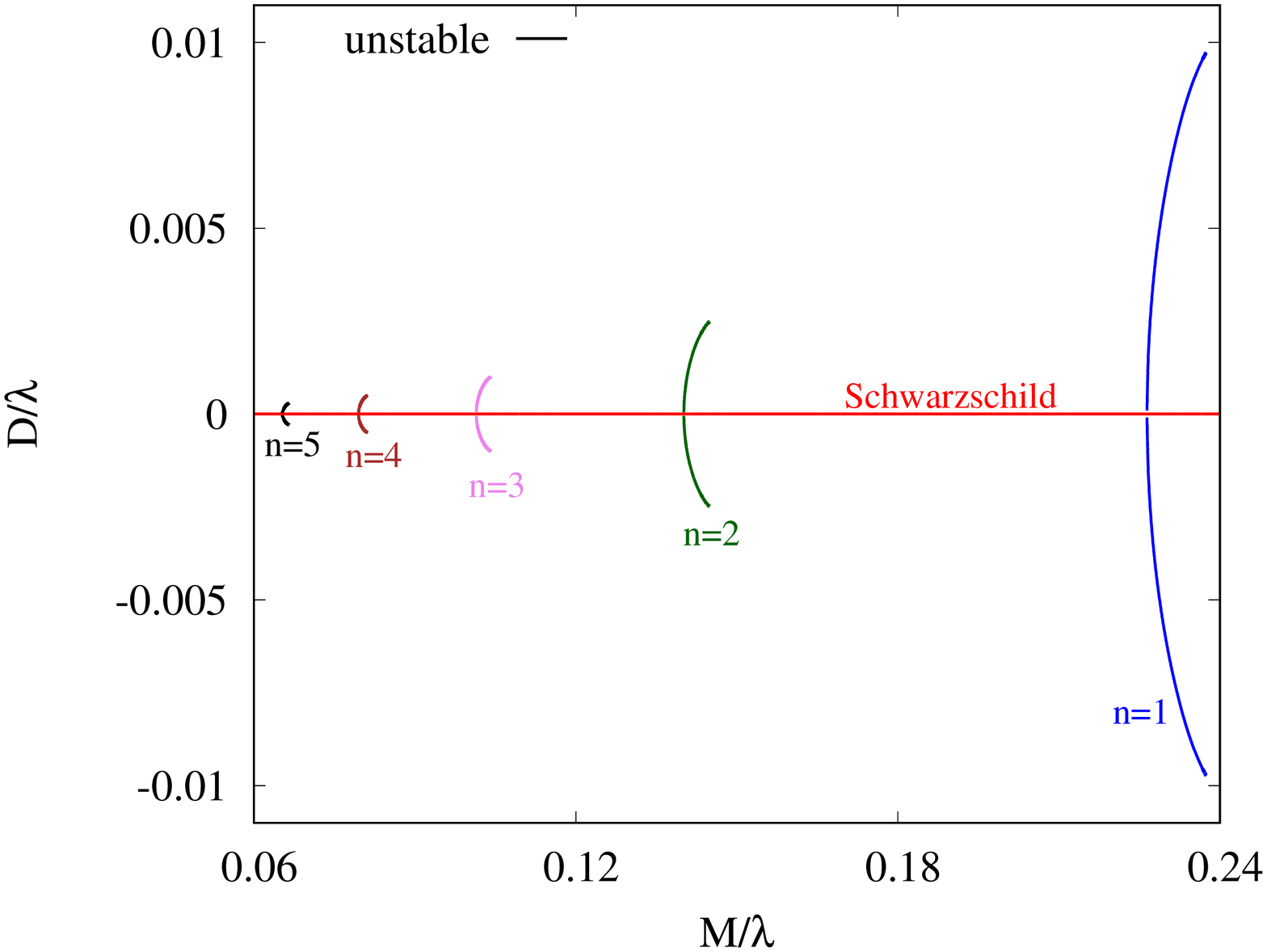}
	\caption{(left) The scalar charge $D$ vs the mass $M$, both quantities scaled with $\lambda$. In red we show the Schwarzschild solution, in orange the fundamental branch and in blue, green, violet, brown and black the branches $n=1, 2, 3 ,4$ and $5$, respectively. Solid curves represent radially unstable solutions, dashed curves solutions without unstable modes, and dotted curves solutions with a strictly positive potential. (right) A zoom of the $n=1...5$ branches. All the solutions shown in this region are unstable.}
	\label{Fig:M_D}
\end{figure}

The first few branches can also be seen in Fig.~\ref{Fig:M_D}, where we exhibit the scalar charge $D/\lambda$ vs the mass $M/\lambda$. The Schwarzschild solution exists for arbitrary values of $M/\lambda$, while the scalarized branches exist only in certain intervals. The fundamental branch (shown in orange) exists between $M/\lambda=0$ and the upper bound $M/\lambda=0.587$, where the scalar hair disappears and the branch merges with the Schwarzschild branch.

The radially excited branches exist only in very small intervals of $M/\lambda$ {since the condition \eqref{eq:BC_sqrt_rh} is quickly violated}. All these branches have the same structure, as seen in Figure \ref{Fig:M_D}. The $n \ge 1$ branches bifurcate from the Schwarzschild branch at certain values of $M/\lambda$, which decrease with increasing node number. Along these scalarized branches, the scalar charge increases with increasing $M$. The branches then end at some critical value of $M/\lambda$, where the scalarized solution becomes singular \cite{Doneva:2017bvd}. The branches rapidly decrease in size as the node number increases.

Figure \ref{Fig:M_D} also indicates the radial stability of each branch, discussed in detail below. Solid curves correspond to radially unstable solutions (i.e., on solid curves the solutions possess at least one unstable radial mode). The dotted curves correspond to solutions that are radially stable (i.e., the effective potential is everywhere positive). The dashed curves correspond to solutions that do not seem to possess unstable modes although the potential is not strictly positive.

\begin{figure}[]
	\centering
	\includegraphics[width=0.38\textwidth,angle=-90]{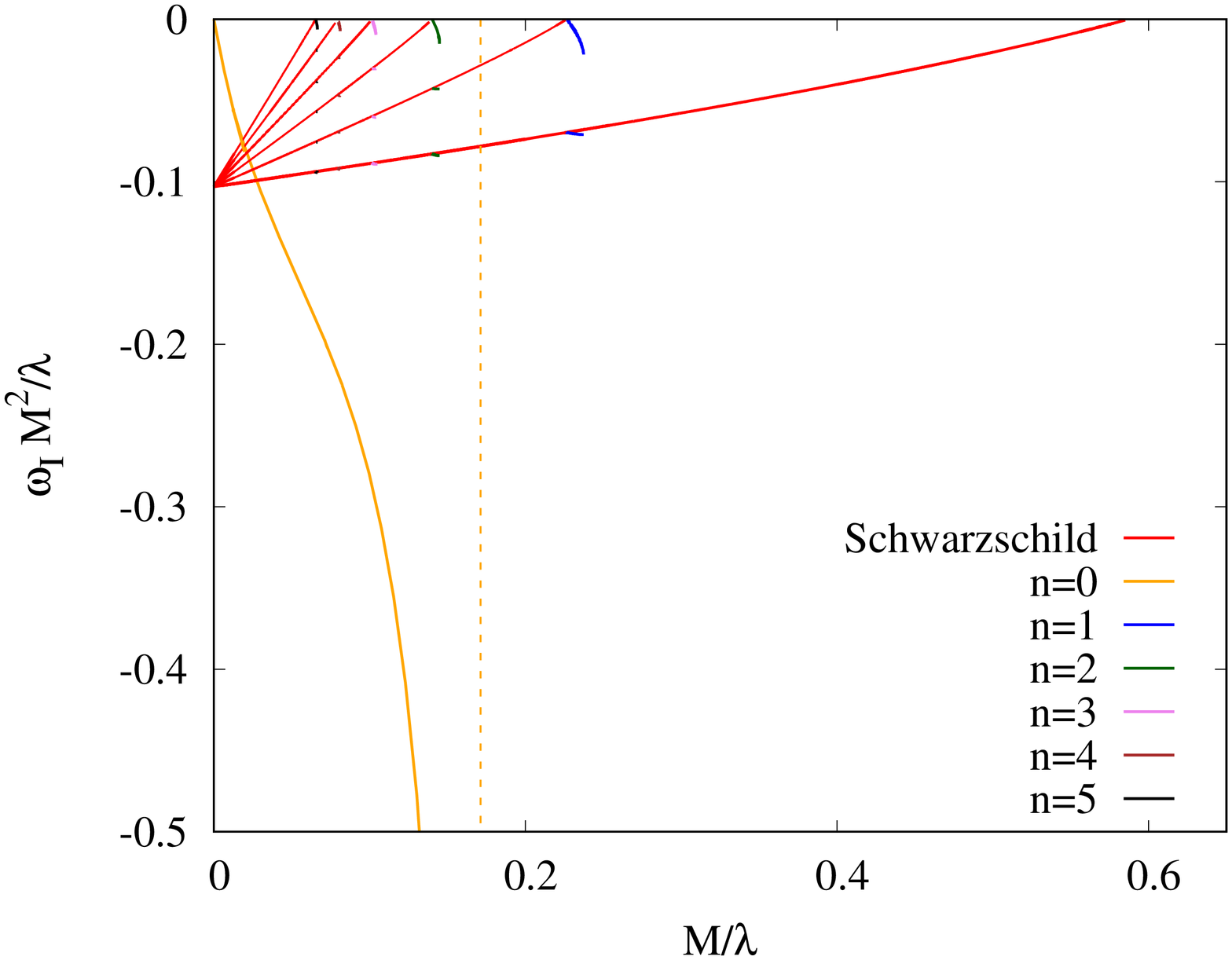}
	\includegraphics[width=0.38\textwidth,angle=-90]{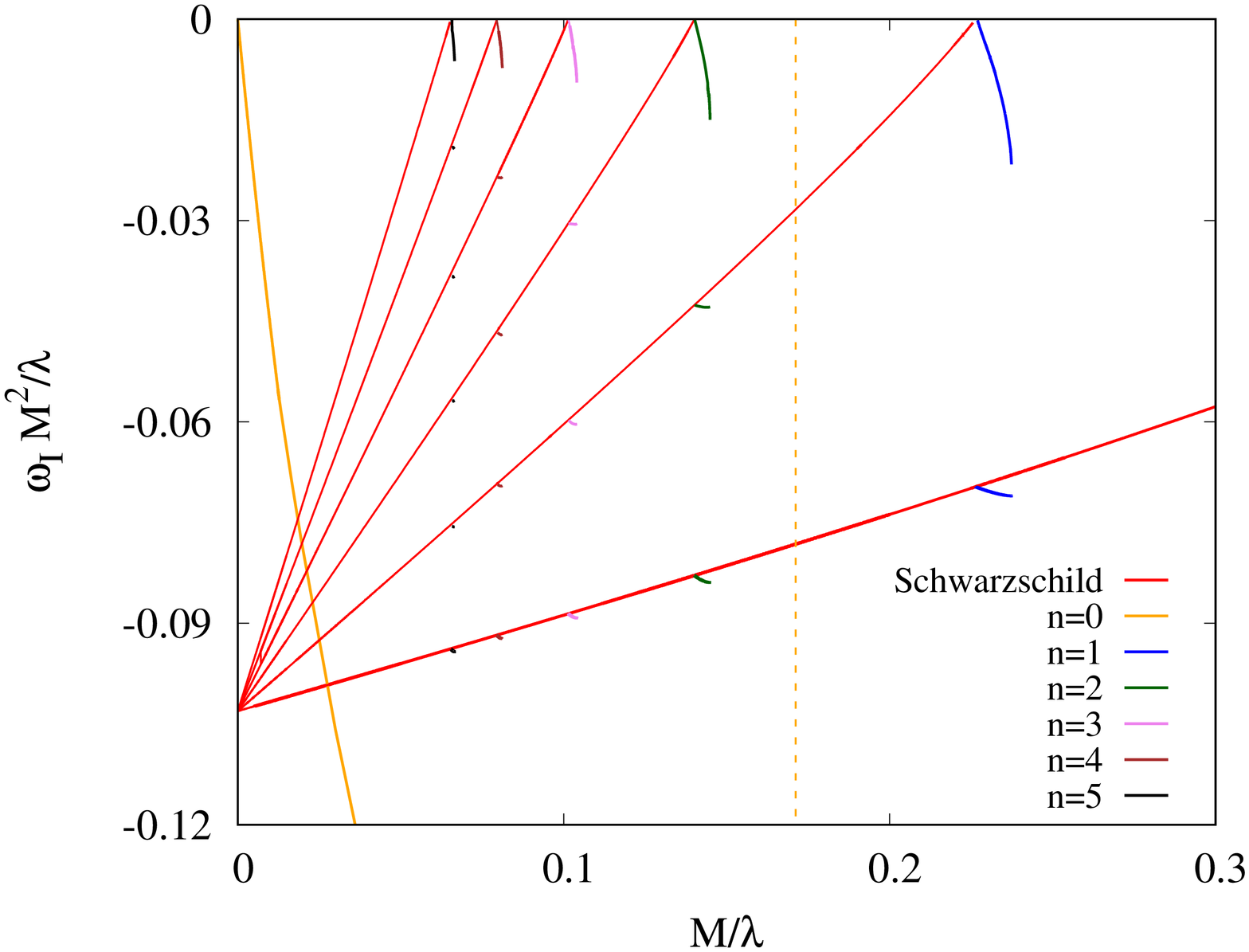}
	\caption{(left) Eigenfrequency $\omega_I$ scaled with $M^2/\lambda$ vs the scaled mass $M/\lambda$. The unstable modes in red correspond to the Schwarzschild solution. Note, that there are no unstable Schwarzschild modes beyond $M/\lambda=0.587$. In orange we show the mode corresponding to the fundamental branch. The vertical dashed line in orange marks the value $M/\lambda=0.171$ at which the unstable mode of the fundamental branch disappears. In blue, green, pink, brown and black we show the modes corresponding to the $n=1...5$ branches, respectively. (right) A zoom focused on the modes of the $n=1...5$ branches. 
 }
	\label{Fig:branches_modes_scaled}
\end{figure}

\subsubsection{Unstable Schwarzschild modes}

Let us now turn to a detailed discussion of the radial stability of the Schwarzschild branch and the scalarized branches, starting with a summary of our findings as exhibited in Fig.~\ref{Fig:branches_modes_scaled}. Here we exhibit the value of $\omega_I$ multiplied by $M^2/\lambda$ vs the (scaled) mass $M/\lambda$ for all considered branches of solutions. In red we show the modes of the Schwarzschild solution, and with various other colours the modes corresponding to the scalarized branches. The right panel of Fig.~\ref{Fig:branches_modes_scaled} shows a zoom focusing on the region, where the $n=1...5$ scalarized branches reside.

Fig.~\ref{Fig:branches_modes_scaled} shows that the branch of Schwarzschild solutions possesses sets of unstable modes, indicated by the red curves. All these modes behave like $(M^2/\lambda)^{-1}$ when $M/\lambda \to 0$ (see the scaling). The first set of unstable modes extends from $M/\lambda=0$ up to $M/\lambda=0.587$. This set of modes corresponds to the instability previously investigated in \cite{Doneva:2017bvd} by analyzing the potential. At $M/\lambda=0.587$ the curve reaches a zero mode (i.e., $\omega_I=0$), which appears because at that point there is the bifurcation point of the fundamental branch. 
This set of unstable modes exists in the same range of parameters as the fundamental branch, which coexists with the Schwarzschild solution in this region of the parameter space (see Fig.~\ref{Fig:M_D} (left), the orange curve). 

Consequently the Schwarzschild black hole in this theory is unstable under radial perturbations in the full intervall where the fundamental branch of scalarized black holes exists. Moreover, the black hole solutions are not unique in this interval. Nevertheless, as soon as the fundamental branch ceases to exist (i.e., for $M/\lambda>0.587$), the instability of the Schwarzschild solutions disappears. 
The potential, however, is only strictly positive for $M/\lambda>\sqrt{3/4}$.
Hence in Fig.~\ref{Fig:M_D} (left) a solid red line indicates the unstable Schwarzschild solutions in the interval $0< M/\lambda < 0.587$, a dashed line indicates the solutions between $0.587<M/\lambda<\sqrt{3/4}$, and a dotted line marks the stable Schwarzschild solutions for $0.587 < M/\lambda$.

Clearly, this is not the only set of unstable modes that the Schwarzschild solution possesses. In Fig.~\ref{Fig:branches_modes_scaled} we see additional sets of unstable modes extending from $M/\lambda=0$ up to certain values of $M/\lambda$, where further zero modes are reached. Again, these zero modes appear precisely at the bifurcation points of the scalarized branches, but now with node number $n>0$ (shown by the short blue to black curves in Fig.~\ref{Fig:M_D}). We note, that the different sets of unstable modes of the Schwarzschild solutions can be characterized by the number of nodes of the perturbation function $\varphi_1$: the set of unstable Schwarzschild modes connected to the zero mode associated with the emergence of the $n$-th scalarized branch always possesses $n$ nodes in the function $\varphi_1$.

%

To summarize the above analysis, we conclude that {\sl the number of unstable modes of the Schwarzschild solution depends on the value of $M/\lambda$: if the Schwarzschild solution resides in between bifurcation points of the $n$-th and the $(n+1)$-th branch of scalarized solutions, then it has $n+1$ unstable modes.} 

\subsubsection{Unstable $n>0$ modes}

Let us next address the unstable modes of the $n>0$ branches before discussing the stability of the fundamental branch. 
In Fig.~\ref{Fig:branches_modes_scaled} these unstable modes are marked in blue, green, pink, brown and black for the branches with $n=1,2,3,4$ and $5$, respectively. We see that these scalarized black holes also possess several sets of radially unstable modes.
One of these sets is always related to the respective zero mode of the Schwarzschild solution at the bifurcation point. For instance, in Fig.~\ref{Fig:branches_modes_scaled} we see that at each zero mode (for $n \ge 1$), a set of unstable modes of the corresponding $n$-th branch appears and extends up to the maximum value of $M/\lambda$, where the background solutions become singular.

\begin{figure}[]
	\centering
	\includegraphics[width=0.38\textwidth,angle=-90]{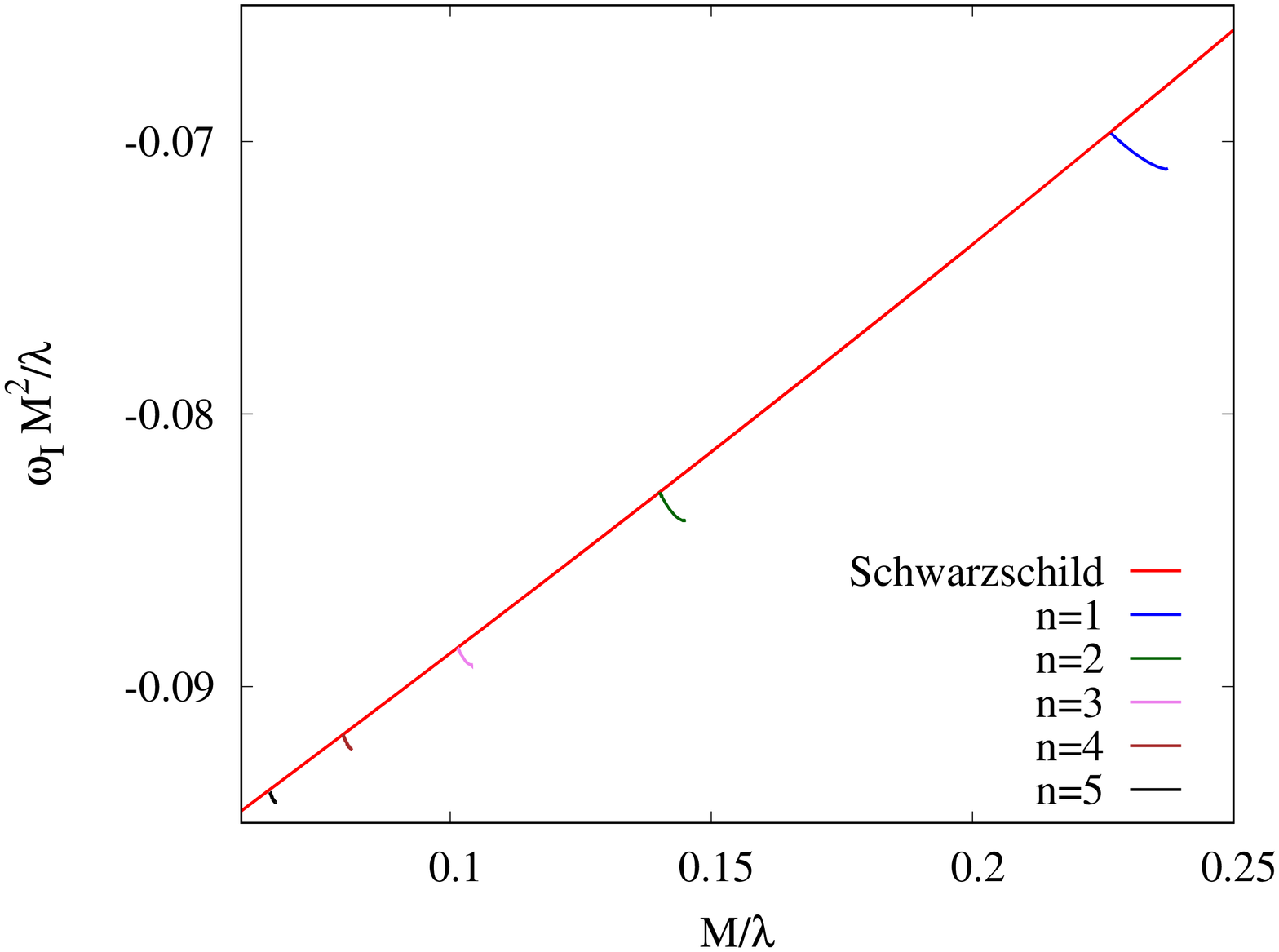}
	\includegraphics[width=0.38\textwidth,angle=-90]{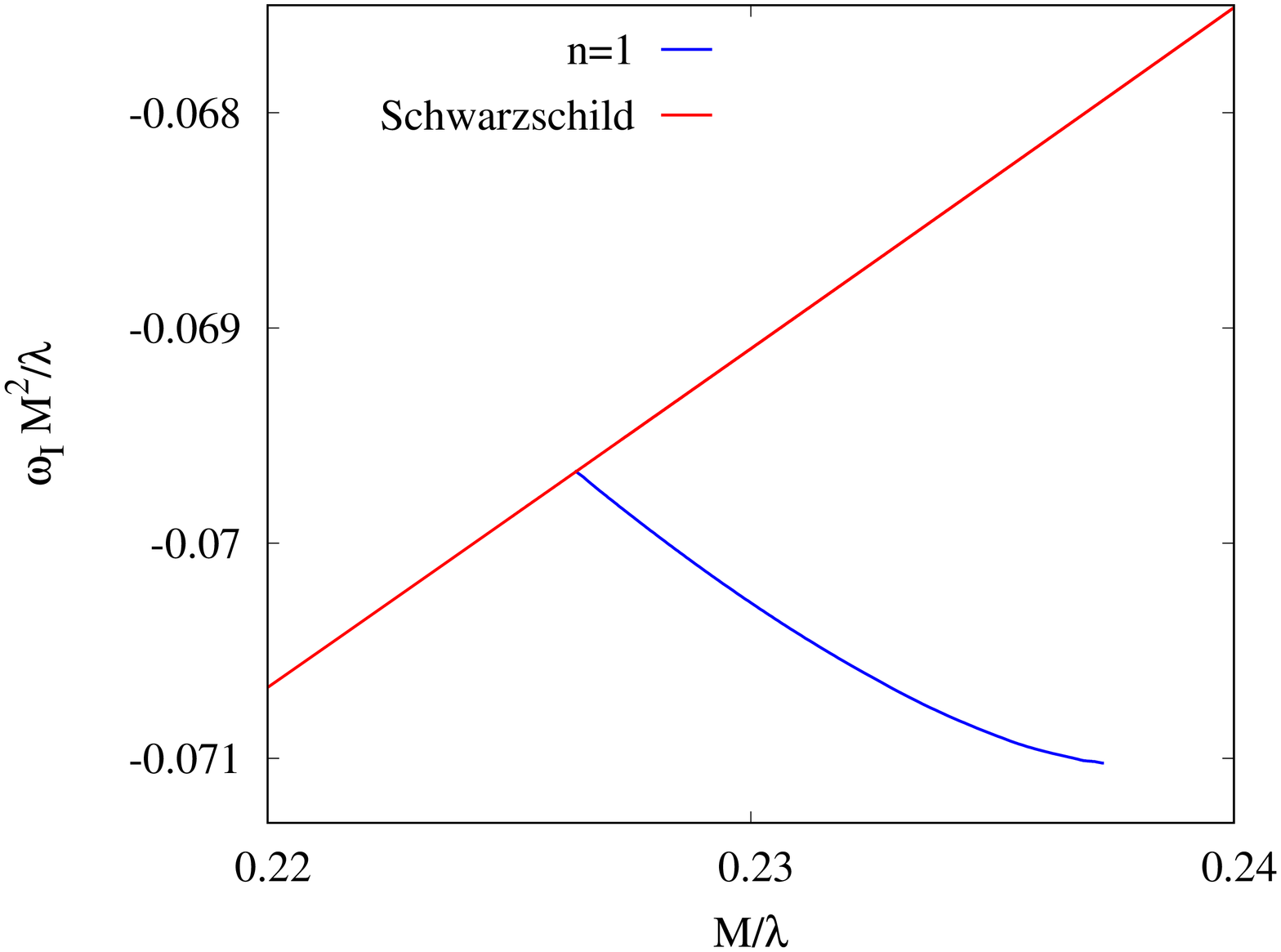}
	\caption{(left) Subset of eigenfrequencies $\omega_I$ scaled with $M^2/\lambda$ vs the scaled mass $M/\lambda$. The unstable modes in red correspond to those of the Schwarzschild solution and connect with the zero mode associated with the bifurcation point of the fundamental branch. In further colors shown are a subset of unstable modes of the $n>0$ branches. (right) A zoom focusing on a set of unstable modes of the $n=1$ branch (blue).}
	\label{Fig:branches2}
\end{figure}

The additional sets of unstable modes are not directly connected to the zero modes. 
However, since the $n$-th excited branch bifurcates from the Schwarzschild solution, 
its unstable modes bifurcate from the unstable modes 
of the Schwarzschild solution
as well as from the zero mode.
This occurs
at the bifurcation point as dictated by continuity 
(although for zero modes there could also arise a mode that is not 
purely imaginary). 
Therefore the $n$ excited branch should have a total of $n+1$ unstable modes, as indeed seen in Fig.~\ref{Fig:branches_modes_scaled}.
In Fig.~\ref{Fig:branches2} (left) we focus on the subset of unstable modes that bifurcate from the set of unstable Schwarzschild modes connecting with the zero mode of the fundamental branch, while in Fig.~\ref{Fig:branches2} (right) we zoom in on these unstable modes of the $n=1$ branch. The behaviour is analogous for the sets of unstable modes bifurcating from the other branches of unstable Schwarzschild modes.
We also note, that each set of unstable modes can be characterized by the nodes in the $\varphi_1$ function, possessing the same number of nodes as the modes of the corresponding Schwarzschild family they are connected with.

Let us again summarize the above analysis and conclude that {\sl the $n$-th excited branch of scalarized black holes (where $n>0$) possesses $n+1$ distinct unstable modes.}

\begin{figure}[]
	\centering
	\includegraphics[width=0.38\textwidth,angle=-90]{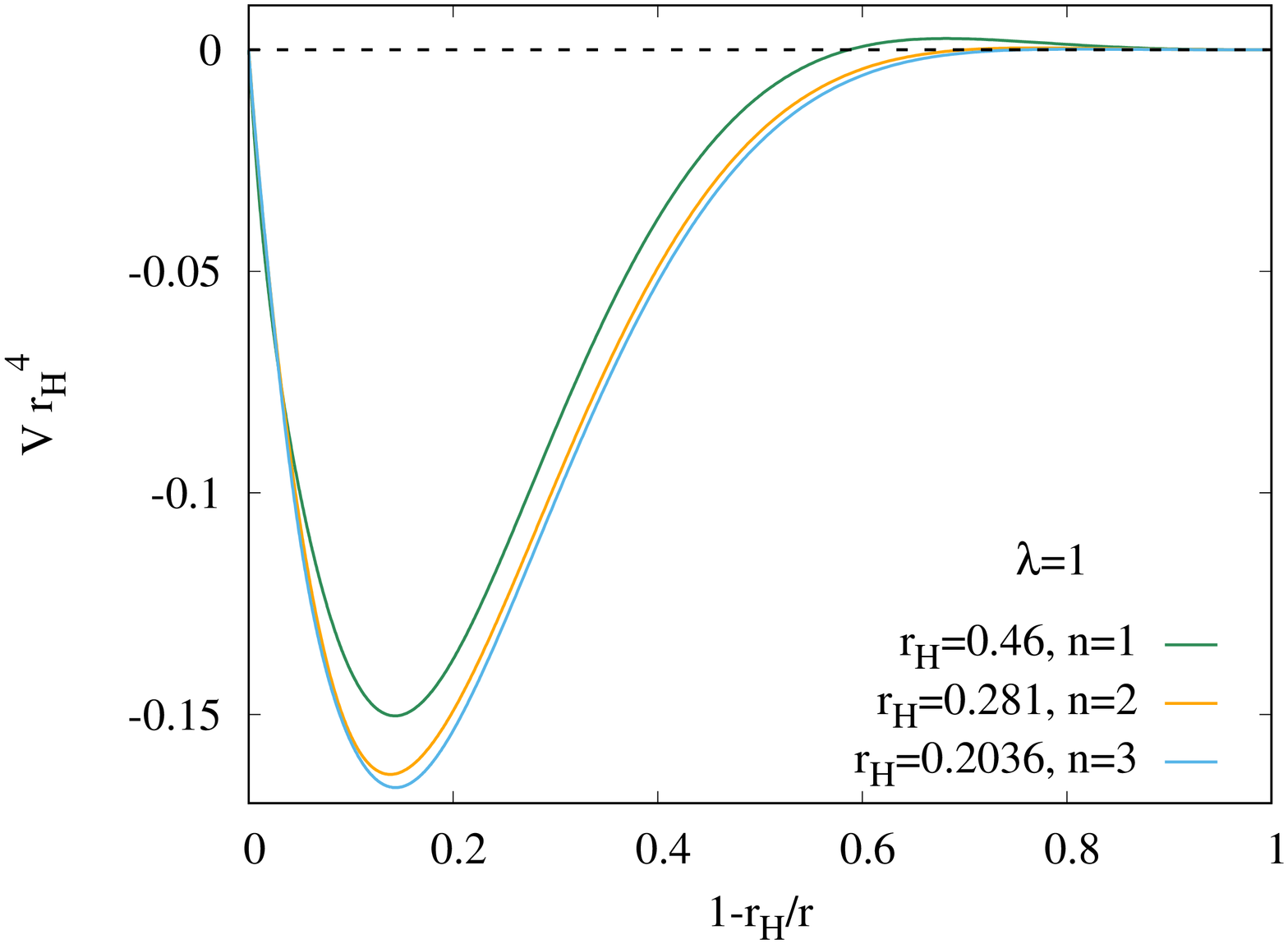}
	\includegraphics[width=0.38\textwidth,angle=-90]{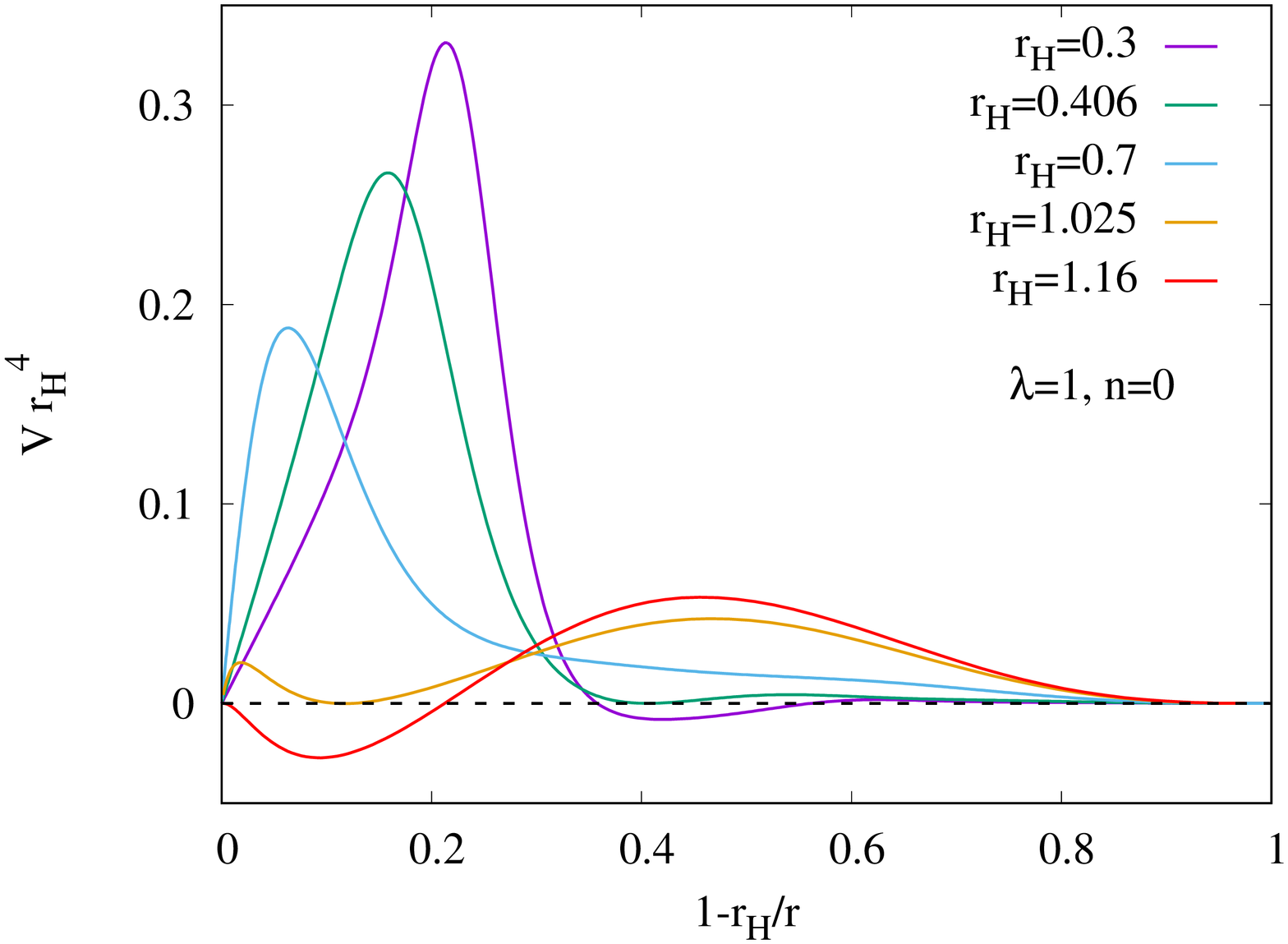}
	\caption{(left) The scaled potential $V$ vs $1-r_H/r$ for several black holes on the excited branches with $n=1$, 2 and 3 for $\lambda=1$ and several values of $r_H$. (right) A similar figure for black holes on the fundamental branch.}
	\label{Fig:potential}
\end{figure}

This conclusion is supported further by an inspection of the potential for the excited branches. In Fig.~\ref{Fig:potential} (left) we show the potential $V$ vs the function of the radial coordinate $1-\frac{r_H}{r}$ for black holes of the $n=1$, $2$ and $3$ branches. Here we clearly see that the potential is dominantly negative. In fact, the integral of the potential is negative
\begin{eqnarray}
n>0 \ \Rightarrow \ \ \int^{\infty}_{-\infty} V(R) dR < 0 \ ,
\end{eqnarray}
in accordance with the existence of the unstable modes.
Hence in Fig.~\ref{Fig:M_D} (left), the $n>1$ branches are shown by solid lines, since they are always radially unstable.

\subsubsection{Stability analysis of the fundamental branch}

Let us now turn to the analysis of the radial stability of the fundamental branch of scalarized black holes. 
In Fig.~\ref{Fig:potential} (right) we show the potential $V$ vs the function of the radial coordinate $1-\frac{r_H}{r}$ for several black holes along this branch (with $\lambda=1$ and different values of $r_H$). What we observe is that the potential is always positive for solutions in the range $0.285 \lesssim M/\lambda \lesssim  0.542$. Hence we conclude that no unstable modes exist on the fundamental branch for black holes as long as the solution belongs to this region. In Fig.~\ref{Fig:M_D} (left) this region of the fundamental branch (orange) is marked with a dotted line. 

For black holes on the fundamental branch in the interval $0.542 \lesssim M/\lambda \lesssim 0.587$ the potential becomes slightly negative in a region close to the horizon, while for black holes with $0.171\lesssim M/\lambda\lesssim 0.288$ the potential becomes slightly negative in some intermediate region of $r$,
as seen, for instance, in Fig.~\ref{Fig:potential} (left) for the potentials corresponding to $\lambda=1$, $r_H=1.16$ (in red) and $r_H=0.3$ (in purple).
However for both intervals the integral of the potential is always positive, meaning that
\begin{eqnarray}
n=0, \ M/\lambda \gtrsim 0.171 \ \Rightarrow \ \ \int^{\infty}_{-\infty} V(R) dR > 0 \ .
\end{eqnarray}
Although this does not exclude the possibility of unstable modes,  we have not been able to generate any solution to the perturbation equation (\ref{master_eq}) describing an unstable mode and satisfying the boundary and regularity conditions. We interpret this as a strong indication that solutions along this branch are mode stable as long as $M/\lambda \gtrsim 0.171$. In Fig.~\ref{Fig:M_D} (left) the two intervals of the fundamental branch are marked with a dashed line, where stability can thus not be decided with certainty.

\begin{figure}[]
	\centering
	\includegraphics[width=0.38\textwidth,angle=-90]{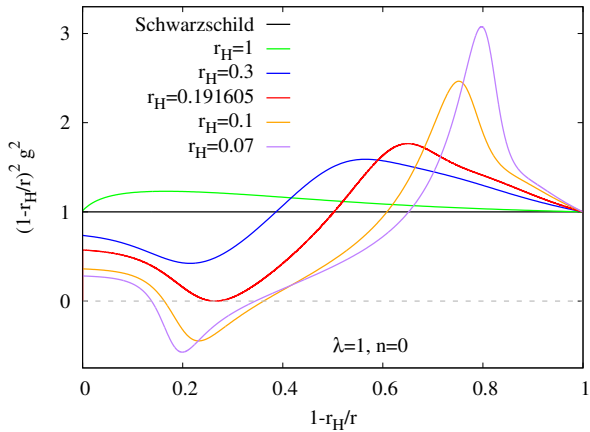}
	\caption{The function $(1-\frac{r_H}{r})^2 g^2$ vs $1-r_H/r$ for black holes on the fundamental branch for $\lambda=1$ and several values of $r_H$.}
	\label{Fig:tortoise}
\end{figure}

The black holes on the fundamental branch in the interval $0 < M/\lambda\lesssim 0.171$ need a more involved investigation. Let us  first consider the  problem on the level of the 
Schr\"odinger equation (\ref{master_eq_simp}).
For the interval under consideration the tortoise coordinate becomes ill-defined. In order to show this, we plot in Fig.~\ref{Fig:tortoise} the function $(1-\frac{r_H}{r})^2 g^2$ vs $1-\frac{r_H}{r}$, for black holes belonging to the fundamental branch with $\lambda=1$ and for different values of $r_H$. This function should be positive in order to have a well defined tortoise coordinate $R$ (see eq.~(\ref{eq:g})). As $r_H$ is decreased, the function deviates more and more from the Schwarzschild case, which corresponds to $g=(1-\frac{r_H}{r})^{-1}$. For small enough values of $r_H$, the function becomes negative in the range  $r_H\le r\le r_*$ where $g(r_*)=0$. As a consequence, the potential becomes singular for $M/\lambda\lesssim 0.171$. 

The more fundamental reason  for this strange behavior is that for $M/\lambda\lesssim 0.171$   {equation (\ref{wave_eq})} is not  hyperbolic for $r_H\le r\le r_*$. This is an interesting and highly nontrivial phenomenon and we leave its investigation to
future work. What is immediately clear from this fact is that the study of the linear stability as a Cauchy problem is ill-posed.
Nevertheless, we can formally study the stability on the level of the eigenvalue problem. It is possible to analyze the mode stability by using the standard coordinate $r$ and integrating eq.~(\ref{master_eq}). Then it is possible to find that indeed, black holes of the fundamental branch with $M/\lambda\lesssim 0.171$ possess an unstable mode. In Fig.~\ref{Fig:branches_modes_scaled}, this unstable mode is shown in orange. The mode extends from $M/\lambda=0$ up to $M/\lambda\approx 0.171$, where the mode diverges. (This limit is marked by a vertical dashed line in orange in Fig.~\ref{Fig:branches_modes_scaled} (left).) Interestingly, this unstable mode does not bifurcate from any Schwarzschild mode (although in the Figure it crosses the sets of unstable Schwarzschild modes for small values of $M/\lambda$, it is not connected with them). The perturbation function $\varphi_1$ of this family of modes always presents zero nodes.

Marking consequently this part of the fundamental branch extending from $M/\lambda=0$ to $M/\lambda=0.171$ with a solid orange line in Fig.~\ref{Fig:M_D} we note, that {\sl black holes belonging to the fundamental branch are not always radially stable}, i.e., for arbitrary values of $M/\lambda$ (despite them having always the larger value of the entropy, as it shown in \cite{Doneva:2017bvd}), and even worse, for small enough black holes (as compared to the coupling parameter) the theory is not hyperbolic. Stability is only found for sufficiently large values of $M/\lambda$. 

We remark that the structure found is to some extent reminiscent (although much more complicated) to the radial instability observed in the dilatonic EGB black holes in \cite{Torii:1998gm,Blazquez-Salcedo:2017txk}. This theory corresponds to $f(\varphi)=e^{2\gamma\varphi}$ and $\lambda^2=\frac{\alpha}{4}$ (following the conventions of \cite{Blazquez-Salcedo:2016enn,Blazquez-Salcedo:2017txk}). For certain values of the coupling constant $\gamma$, one secondary branch of solutions is present in a small region of the parameter space, in addition to the main branch of dilatonic black holes, resulting in non-uniqueness of the solutions. The secondary branch appears close to the minimum value of the black hole mass allowed by the theory, and it was found to be radially unstable \cite{Torii:1998gm,Blazquez-Salcedo:2017txk}. 

It is also interesting to note that in dilatonic EGB black holes the existence of a minimum mass is caused by the existence of a limiting value of the normalized coupling constant (i.e., the maximum value of $\zeta=\alpha/M^2$ \cite{Blazquez-Salcedo:2017txk}, when condition (\ref{eq:BC_sqrt_rh}) is no longer satisfied). No regular black hole solutions can be found for smaller values of this mass. 
This is different from the scalarized EGB black holes on the fundamental branch considered here, which exist for arbitrarily small values of the mass. However, this branch also possesses an effective minimum mass ($M/\lambda \approx 0.171$), below which 
the theory is no longer hyperbolic (i.e., where no longer stable configurations can be found).

\subsection{Quadratic coupling function}

\begin{figure}[]
	\centering
	\includegraphics[width=0.38\textwidth,angle=-90]{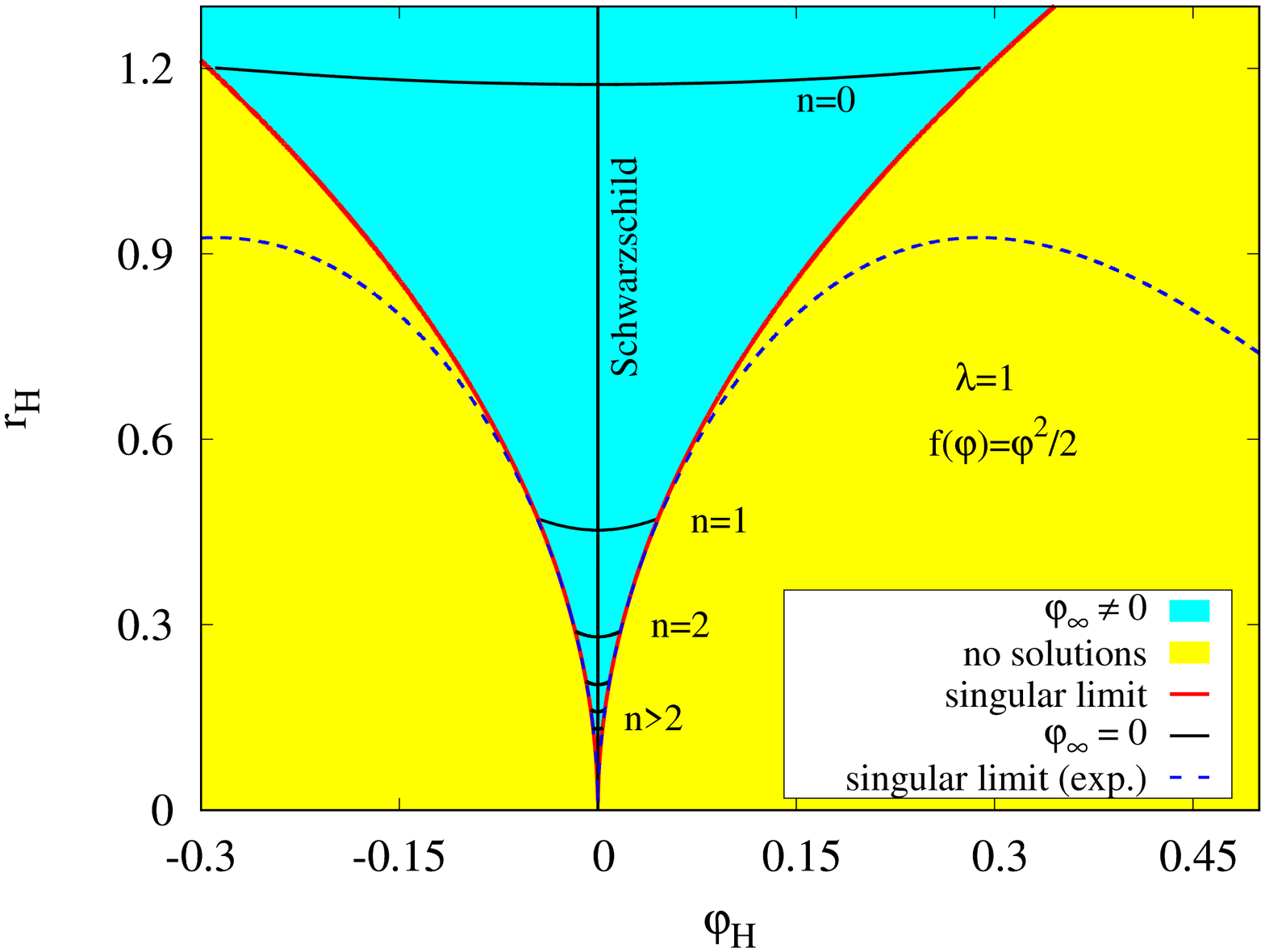}
	\includegraphics[width=0.38\textwidth,angle=-90]{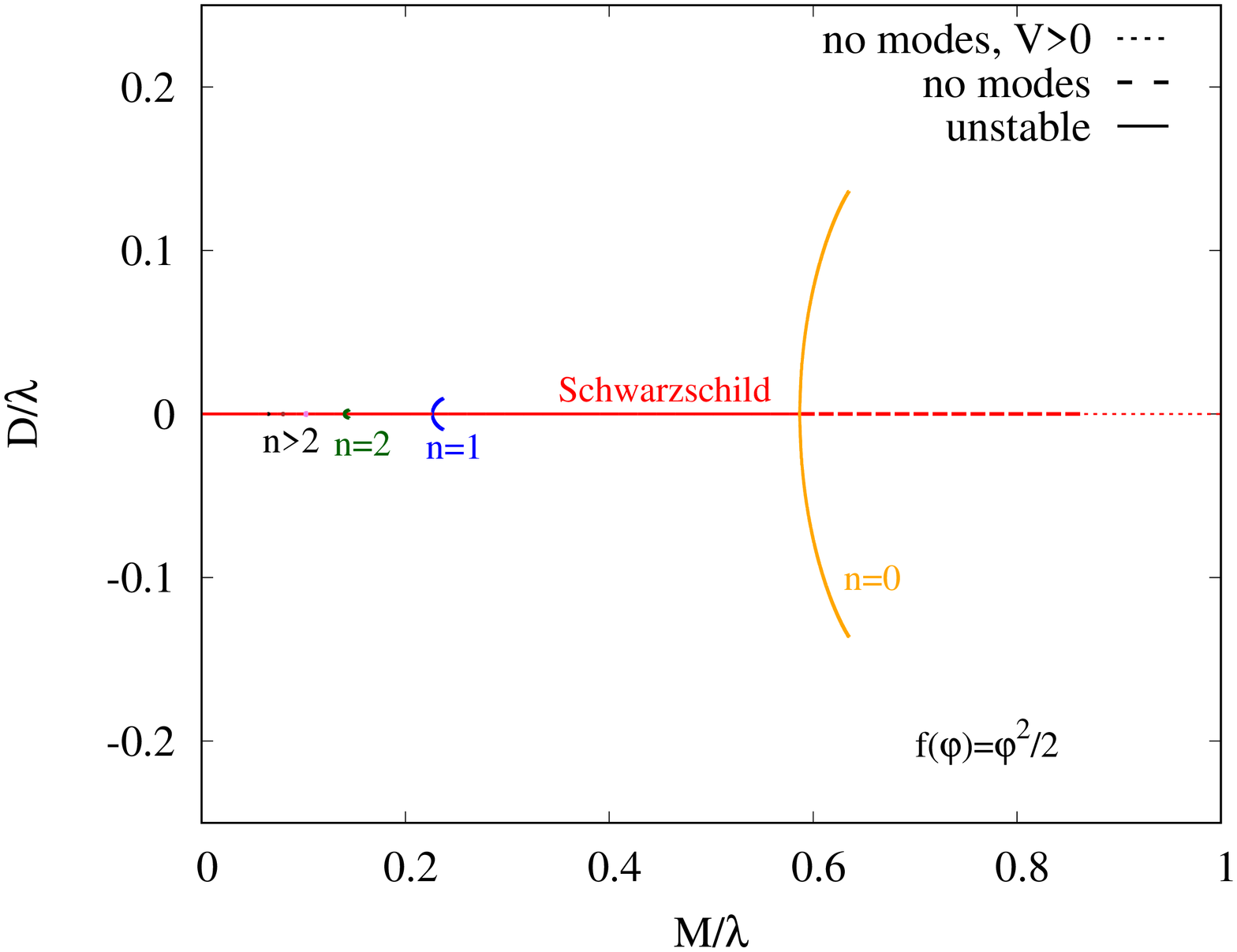}
	\caption{(left) Domain of existence $r_H$ vs $\varphi_H$ of the solutions with the quadratic coupling (\ref{eq:coupling_function_quad}). For comparison, we include the singular limit for the coupling (\ref{eq:coupling_function}) as a dashed blue curve. (right) The scalar charge $D$ vs the mass $M$ for both quantities scaled with $\lambda$.}
	\label{Fig:branches_domain_quad}
\end{figure}

Finally we turn to the case of the quadratic coupling (\ref{eq:coupling_function_quad}). In Fig.~\ref{Fig:branches_domain_quad} (left) we show the domain of existence $r_H$ vs $\varphi_H$. As expected, we find that the space of solutions is very similar to the one of the previous coupling in the small $\varphi_H$ and $r_H$ region (compare Fig.~\ref{Fig:domain}). In particular, since the quadratic coupling (\ref{eq:coupling_function_quad}) is obtained in the small $\varphi$ limit of the exponential coupling (\ref{eq:coupling_function}), the bifurcation points of the Schwarzschild solution coincide. Even more, the structure of the zero modes and sets of instabilities of the Schwarzschild solution is exactly the same for both couplings. This is seen in Fig.~\ref{Fig:branches_domain_quad} (right), where we show $D/\lambda$ vs $M/\lambda$ for solutions with the quadratic coupling.

\begin{figure}[]
	\centering
	\includegraphics[width=0.38\textwidth,angle=-90]{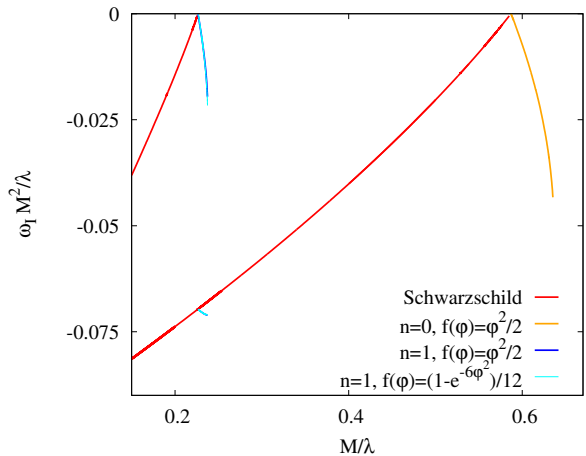}
	\caption{The scaled eigenfrequencies $\omega_I$ vs the scaled mass $M/\lambda$ focused on the $n=0$, $1$ branches for the quadratic coupling (\ref{eq:coupling_function_quad}). For comparison, we include the unstable modes of the $n=1$ branch of the exponential coupling (\ref{eq:coupling_function}) with a cyan curve.}
	\label{Fig:modes_quad}
\end{figure}

The most important difference appears for the fundamental branch $n=0$, which in this case extends from the bifurcation point $M/\lambda=0.587$ to larger values of $M/\lambda$ and reaches the singular limit at a finite value of $\varphi$. This gives the $n=0$ branch a similar structure to the excited $n>0$ branches. In fact, the fundamental branch turns out to be unstable like the excited branches. In Fig.~\ref{Fig:modes_quad} we exhibit the scaled unstable modes $\omega_I$ vs $M/\lambda$ focusing on the $n=0$, $1$ branches for the quadratic coupling. Again in red we show the first two sets of unstable Schwarzschild modes that connect with the $n=0$ and $n=1$ zero modes. In the present case at the zero mode of the $n=0$ branch, a new set of unstable modes appears and remains present on the whole fundamental branch (shown by the orange curve). The branch and with it its set of unstable modes end when the singular configuration is reached. The $n=1$ modes are also shown (in blue). Moreover, for comparison the respective unstable modes corresponding to the exponential coupling (\ref{eq:coupling_function}) are shown (in cyan).

Hence in this case, {\sl all the scalarized branches are unstable, and the $n$-th branch possesses $n+1$ unstable modes, including the fundamental branch, $n=0$.} This means there are no stable configurations below the first branching point at $M/\lambda=0.587$. {As a matter of fact this conclusion coincides with the thermodynamical results -- after performing a calculation of the black hole entropy similar to \cite{Doneva:2017bvd} but for the particular coupling function (\ref{eq:coupling_function_quad}), it turns out that all of the branches with a nontrivial scalar field, including the fundamental one, have lower entropy than the Schwarzschild solution and therefore they are thermodynamically less stable.} 


\section{Conclusions}

In the present paper we have studied the stability with respect to radial perturbations of scalarized black holes in EGB theories. The two particular cases of the coupling functions between the scalar field and the Gauss-Bonnet invariant considered have been motivated by the previous studies of black holes with a nontrivial scalar field. The first one is the case of $f(\varphi)=  \frac{1}{12} \left(1- e^{-6\varphi^2}\right)$ considered in \cite{Doneva:2017bvd} which leads to a well manifested fundamental branch (i.e., the branch characterized by a scalar field which has no nodes) that deviates significantly from the Schwarzschild black holes. Second, the coupling function $f(\varphi)=  \frac{1}{2} \varphi^2$, considered in \cite{Silva:2017uqg}, has been examined.

It was previously shown \cite{Doneva:2017bvd,Silva:2017uqg} that the Schwarzschild solution is stable only up to the first point of bifurcation where the fundamental branch appears.  Therefore a natural question to ask has been whether another solution with a nontrivial scalar field is stable for masses smaller than the critical mass of the bifurcation. The results from the linear stability analysis performed show that all solutions characterized by a scalar field with $n>0$ nodes possess $n+1$ unstable modes. Thus all radially excited branches  of solutions are radially unstable. 

The picture is more complicated for the fundamental branch. In the case of the coupling function  $f(\varphi)=  \frac{1}{12} \left(1- e^{-6\varphi^2}\right)$, the scalarized solutions are stable from the bifurcation point until some small critical mass $M_*$, where they formally lose stability as well. The presence of this instability is formally different in nature from the instability of the other scalarized branches -- the unstable modes are not connected with any zero mode of the Schwarzschild solution and are due most probably to the fact that for small masses the tortoise coordinate is ill defined and leads to singularities of the potential of the perturbations equations. In fact the situation is even worse because below the critical mass $M_*$ the theory loses its hyperbolicity, and the stability has only been investigated formally on the basis of a formal eigenvalue problem.      

In constrast, in the case of the coupling function $f(\varphi)=  \frac{1}{2} \varphi^2$ the whole fundamental branch, which in this case is short and  terminates at some nonzero mass because of violation of condition (\ref{eq:BC_sqrt_rh}), is unstable. This instability is a more ``classical'' one since it is connected with a zero mode of the Schwarzschild solution.

We should note that these conclusions are in agreement with the thermodynamical studies of the stability performed in \cite{Doneva:2017bvd}. More precisely, the entropy of all branches of solutions possessing a scalar field which has one or more nodes is smaller than the entropy of the Schwarzschild black holes. On the other hand, the results show that the fundamental branch for $f(\varphi)=  \frac{1}{12} \left(1- e^{-6\varphi^2}\right)$ is thermodynamically more stable as compared to the Schwarzschild solution, while for $f(\varphi)=  \frac{1}{2} \varphi^2$ the entropy of the fundamental branch is always smaller than the Schwarzschild one. 

It is interesting to make a comparison of these solutions with the black holes in the dilatonic EGB theory, where the coupling function has the form $f(\varphi)=  e^{2\gamma\varphi}$. In this case solutions can only exist if they are larger than a certain minimum value of the mass. For some values of the coupling constant $\gamma$, a secondary branch of black holes appears close to this limit. The solutions on this secondary branch were shown to be always radially unstable \cite{Torii:1998gm,Blazquez-Salcedo:2017txk}.

Thus, the stability and the existence of solutions is highly controlled by the coupling function. A general conclusion, though, is that for all of the considered cases there exists a threshold mass below which there are no stable black hole solutions  (including the Schwarzschild one). As a matter of fact one might be able to cure this problem by a better choice of the coupling function (or even by varying the numerical constants in the coupling functions (\ref{eq:coupling_function}) or (\ref{eq:coupling_function_quad})). Answering this question and studying the loss of hyperbolicity for small black holes  require, however, a much more thorough investigation  that will be the subject of a future study.

\section*{Acknowledgements}
JLBS would like to acknowledge support from the DFG project BL 1553. JLBS and JK would like to acknowledge support by the 
DFG Research Training Group 1620 {\sl Models of Gravity} 
and the COST Action CA16104.
SY and DD would like to thank for support 
by the COST Action CA16214. SY would like to thank for support 
by the  COST Action CA16104 and the Sofia University Research Grant under No 3258. DD would like to thank the European Social Fund, the Ministry of Science, Research and the Arts Baden-W\"urttemberg for the support. DD is indebted to the Baden-W\"urttemberg Stiftung for the financial support of this
research project by the Eliteprogramme for Postdocs. 

\appendix
\section{Additional equations}\label{appendix1}

The functions that appear in the master equation (\ref{master_eq}) depend only on the background configurations. The function $g$ that characterizes the tortoise coordinate can be written like
\begin{eqnarray}
g^2 = A/B \ ,
\end{eqnarray}
with the functions $A$ and $B$:
\begin{eqnarray}
A&=&-8 {{e}^{6\Lambda}}{\varphi_0'}{\lambda}^{2}{r}^{3}{\frac{df}{d\varphi_0}}+{{e}^{8\Lambda}}{r}^{4}
+16{{\varphi_0'}}^{2}{\lambda}^{6} \left( {{e}^{2\Lambda}}-1 \right) ^{2} {{e}^{2\Lambda}}\left({{\frac{df}{d\varphi_0}}}\right)^{2}{\frac{d^2f}{d\varphi_0^2}} \nonumber \\ 
&&-16{\lambda}^{6} \left( {{e}^{2\Lambda}}-1 \right)\left({{e}^{2\Lambda}}{\Lambda'}{\varphi_0'}- {{e}^{2\Lambda}}{\varphi_0''}+3{\Lambda'}{\varphi_0'}+{\varphi_0''} \right)  {{e}^{2\Lambda}}\left({{\frac{df}{d\varphi_0}}}\right)^{3} \nonumber \\ 
&&-4 {{e}^{4\Lambda}}{\lambda}^{4} \left(  {{e}^{4\Lambda}}-4 {{e}^{2\Lambda}}{\Lambda'}r-4{{\varphi_0'}}^{2}{r}^{2}-2 {{e}^{2\Lambda}}+4{\Lambda'}r+1 \right) \left({{\frac{df}{d\varphi_0}}}\right)^{2} \ ,
\end{eqnarray}
\begin{eqnarray}
B&=&-8{{e}^{4\Lambda+2\Phi}}{\varphi_0'}{\lambda}^{2}{r}^{3}{\frac{df}{d\varphi_0}}+{{e}^{6\Lambda+2\Phi}}{r}^{4}+16{{e}^{2\Phi}}{\Phi'}{\varphi_0'}{\lambda}^{6} \left( {{e}^{4\Lambda}}+2{{e}^{2\Lambda}}-3 \right) \left({{\frac{df}{d\varphi_0}}}\right)^{3} \nonumber \\ 
&&-4{{e}^{2\Phi+2\Lambda}}{\lambda}^{4} \left( -4{{\varphi_0'}}^{2}{r}^{2}+4{{e}^{2\Lambda}}{\Phi'}r+{{e}^{4\Lambda}}-4{\Phi'}r-2{{e}^{2\Lambda}}+1 \right) \left({{\frac{df}{d\varphi_0}}}\right)^{2} 
\ .
%
\end{eqnarray}

The coefficient $C_1$ can be written like
\begin{eqnarray}
C_1 = C_2/C_4 \ , 
\end{eqnarray}
with the functions $C_2$ and $C_4$ being:
\begin{eqnarray}
C_4 = &&\left[ 2{\frac{df}{d\varphi_0}}\varphi_0' \left({{e}^{2\Lambda}}-3 \right) {\lambda}^{2}+ {{e}^{2\Lambda}}r\right] \times \nonumber \\ 
&&\bigg[ 
 -8{\frac{df}{d\varphi_0}} {{e}^{4\Lambda}}\varphi_0'{\lambda}^{2}{r}^{3}+ {{e}^{6\Lambda}}{r}^{4}+ 16{\Phi'}\varphi_0'{\lambda}^{6} \left( {{e}^{2\Lambda}}-1 \right) \left({{e}^{2\Lambda}}+3 \right) \left({{\frac{df}{d\varphi_0}}}\right)^{3} \nonumber \\ 
 &&-4 {{e}^{2\Lambda}}{\lambda}^{4} \left( 4{\Phi'} {{e}^{2\Lambda}}r + {{e}^{4\Lambda}} -4{\varphi_0'}^{2}{r}^{2}-4{\Phi'}r-2 {{e}^{2\Lambda}}+1 \right) \left({{\frac{df}{d\varphi_0}}}\right)^{2} \bigg]  \ ,
 \end{eqnarray}
\begin{eqnarray} 
C_2 &=&{\frac{d^2f}{d\varphi_0^2}}
\bigg[
4{{\varphi_0'}}^{2}{\lambda}^{2}{r}^{4} \left( {{e}^{8\Lambda}}-{{e}^{6\Lambda}} \right)+ 32{\Phi'}{{\varphi_0'}}^{3}{\lambda}^{8} \left( 15{{e}^{4\Lambda}}-7{{e}^{2\Lambda}}-5{{e}^{6\Lambda}}-3 \right) \left({{\frac{df}{d\varphi_0}}}\right)^{3}
\nonumber \\ 
&&-8{{\varphi_0'}}^{2}{\lambda}^{6}{{e}^{2\Lambda}} \left( 8{{e}^{4\Lambda}}{\Phi'}r+8{{e}^{2\Lambda}}{\Phi'}r-16{\Phi'}r+9{{e}^{4\Lambda}}-9{{e}^{2\Lambda}}-3{{e}^{6\Lambda}}+3 \right) \left({{\frac{df}{d\varphi_0}}}\right)^{2}
\nonumber \\  
&&+4{\varphi_0'}{\lambda}^{4}r{{e}^{4\Lambda}} \left( {{e}^{4\Lambda}}{{\varphi_0'}}^{2}{r}^{2}-6{{e}^{2\Lambda}}{{\varphi_0'}}^{2}{r}^{2}+5{{\varphi_0'}}^{2}{r}^{2}+8{{e}^{2\Lambda}}{\Phi'}r+2{{e}^{4\Lambda}}-8{\Phi'}r-4{{e}^{2\Lambda}}+2 \right) {\frac{df}{d\varphi_0}}
  \bigg]  \nonumber \\ 
&&+32{\varphi_0'}{\lambda}^{8} 
\left({{\frac{df}{d\varphi_0}}}\right)^{4}
\bigg[ 
-7{{e}^{4\Lambda}}{{\Phi'}}^{2}{\varphi_0'}-11{{e}^{4\Lambda}}{\Phi'}{\Lambda'}{\varphi_0'}+15{{e}^{2\Lambda}}{{\Phi'}}^{2}{\varphi_0'}-9{{e}^{2\Lambda}}{\Phi'}{\Lambda'}{\varphi_0'}+{{\Phi'}}^{2}{\varphi_0'}{{e}^{6\Lambda}} 
\nonumber \\ 
&&+9{\Phi'}{\Lambda'}{\varphi_0'}{{e}^{6\Lambda}}+{{e}^{4\Lambda}}{\Phi''}{\varphi_0'}
+13{{e}^{4\Lambda}}{\Phi'}{\varphi_0''}+9{{e}^{2\Lambda}}{\Phi''}{\varphi_0'}-25{{e}^{2\Lambda}}{\Phi'}{\varphi_0''}-{\Phi''}{\varphi_0'}{{e}^{6\Lambda}}
\nonumber \\ 
&&-9{{\Phi'}}^{2}{\varphi_0'}+27{\Phi'}{\Lambda'}{\varphi_0'}-3{\Phi'}{\varphi_0''}{{e}^{6\Lambda}}-9{\Phi''}{\varphi_0'}+15{\Phi'}{\varphi_0''} \bigg]
\nonumber \\ 
&&+8{\lambda}^{6}{{e}^{2\Lambda}}\left({{\frac{df}{d\varphi_0}}}\right)^{3}
\bigg[ 
6{\Lambda'}{\varphi_0'}+18{\Phi''}{\varphi_0'}r+6{\Phi'}{{\varphi_0'}}^{3}{r}^{2}-6{\Lambda'}{{\varphi_0'}}^{3}{r}^{2}+12{{\Phi'}}^{2}{\varphi_0'}r
\nonumber \\
&&+40{{e}^{2\Lambda}}{\Phi'}{\Lambda'}{\varphi_0'}r+3{\varphi_0''}+{{e}^{4\Lambda}}{\varphi_0''}+{\varphi_0''}{{e}^{6\Lambda}}-5{{e}^{2\Lambda}}{\varphi_0''}+24{{\varphi_0'}}^{3}r-6{{e}^{4\Lambda}}{\Phi'}{{\varphi_0'}}^{3}{r}^{2}-2{{e}^{4\Lambda}}{\Lambda'}{{\varphi_0'}}^{3}{r}^{2}
\nonumber \\
&&+8{{e}^{2\Lambda}}{\Phi'}{{\varphi_0'}}^{3}{r}^{2}+16{{e}^{2\Lambda}}{\Lambda'}{{\varphi_0'}}^{3}{r}^{2}+4{{e}^{4\Lambda}}{{\Phi'}}^{2}{\varphi_0'}r-16{{e}^{2\Lambda}}{{\Phi'}}^{2}{\varphi_0'}r+2{{e}^{4\Lambda}}{\Phi''}{\varphi_0'}r-12{{e}^{4\Lambda}}{\Phi'}{\varphi_0''}r
\nonumber \\
&&-20{{e}^{2\Lambda}}{\Phi''}{\varphi_0'}r+32{{e}^{2\Lambda}}{\Phi'}{\varphi_0''}r-6{\Phi'}{\varphi_0'}-8{{e}^{2\Lambda}}{{\varphi_0'}}^{3}r-2{{e}^{4\Lambda}}{\Phi'}{\varphi_0'}+6{{e}^{2\Lambda}}{\Phi'}{\varphi_0'}+2{\Phi'}{\varphi_0'}{{e}^{6\Lambda}}
\nonumber \\
&&-20{\Phi'}{\varphi_0''}r-72{\Phi'}{\Lambda'}{\varphi_0'}r
-6{\Lambda'}{\varphi_0'}{{e}^{6\Lambda}}+10{{e}^{4\Lambda}}{\Lambda'}{\varphi_0'}-10{{e}^{2\Lambda}}{\Lambda'}{\varphi_0'} \bigg] 
\nonumber \\
&&+4{\lambda}^{4}{{e}^{4\Lambda}} \left({{\frac{df}{d\varphi_0}}}\right)^{2} \bigg[ -1-4{\Lambda'}r-34{{\varphi_0'}}^{2}{r}^{2}+6{\Phi'}r
+26{\Phi'}{\Lambda'}{r}^{2}-13{\Phi'}{{\varphi_0'}}^{2}{r}^{3}
\nonumber \\
&&-2{{\Phi'}}^{2}{r}^{2}-4{\Phi''}{r}^{2}
+{{e}^{4\Lambda}}{\Phi'}{{\varphi_0'}}^{2}{r}^{3}-{{e}^{4\Lambda}}{\Lambda'}{{\varphi_0'}}^{2}{r}^{3}+{{e}^{4\Lambda}}{\varphi_0'}{\varphi_0''}{r}^{3}-4{{e}^{2\Lambda}}{\Lambda'}{{\varphi_0'}}^{2}{r}^{3}
-6{{e}^{2\Lambda}}{\varphi_0'}{\varphi_0''}{r}^{3}
\nonumber \\
&&+5{\varphi_0'}{\varphi_0''}{r}^{3}+{\Lambda'}{{\varphi_0'}}^{2}{r}^{3}
+{{e}^{4\Lambda}}+{{e}^{2\Lambda}}+8{{e}^{2\Lambda}}{{\varphi_0'}}^{2}{r}^{2}+4{{e}^{2\Lambda}}{\Lambda'}r+2{{e}^{4\Lambda}}{{\varphi_0'}}^{2}{r}^{2}+2{{e}^{2\Lambda}}{{\Phi'}}^{2}{r}^{2} 
\nonumber \\
&&-18{{e}^{2\Lambda}}{\Phi'}{\Lambda'}{r}^{2}-{{e}^{6\Lambda}}+4{{e}^{2\Lambda}}{\Phi''}{r}^{2}+2{{e}^{4\Lambda}}{\Phi'}r-8{{e}^{2\Lambda}}{\Phi'}r
\bigg] -{{e}^{8\Lambda}}{r}^{4} \bigg[ -{\Lambda'}r+{\Phi'}r+2 \bigg] 
\nonumber \\
&&-2{\lambda}^{2}{r}^{3}{{e}^{6\Lambda}} {\frac{df}{d\varphi_0}} \bigg[ {{e}^{4\Lambda}}{\varphi_0'}-2{{e}^{2\Lambda}}{\varphi_0''}r-6{\Phi'}{\varphi_0'}r+2{\Lambda'}{\varphi_0'}r+2{{e}^{2\Lambda}}{\varphi_0'}+2{\varphi_0''}r-15{\varphi_0'} \bigg] \ .
\end{eqnarray}

Finally, the function $U$, related with the effective potential, can be written like
\begin{eqnarray}
U = C_3/C_4 \ ,
\end{eqnarray}
with
\begin{eqnarray}
C_3=
D_0+D_2\lambda^2+D_4\lambda^4+D_6\lambda^6+D_8\lambda^8 ,
\end{eqnarray}
and
\begin{eqnarray}
D_0=-{\varphi_0'}{r}^{5}{{e}^{8\Lambda}} \left( -2{\Phi'}{\varphi_0'}r+2{\Lambda'}{\varphi_0'}r+{{e}^{2\Lambda}}{\varphi_0'}-3{\varphi_0''}r-5{\varphi_0'} \right) ,
\end{eqnarray}
\begin{eqnarray}
D_2&=&-2{r}^{3}{{e}^{6\Lambda}} 
\bigg[ 
{{e}^{2\Lambda}}{{\varphi_0'}}^{4}{r}^{2}-3{{\varphi_0'}}^{4}{r}^{2}-2{{e}^{2\Lambda}}{\Phi'}{{\varphi_0'}}^{2}r+2{{e}^{2\Lambda}}{\Lambda'}{{\varphi_0'}}^{2}r-3{{e}^{2\Lambda}}{\varphi_0'}{\varphi_0''}r+4{\Phi'}{{\varphi_0'}}^{2}r
\nonumber \\
&&-4{\Lambda'}{{\varphi_0'}}^{2}r+{{e}^{4\Lambda}}{{\varphi_0'}}^{2}-{{e}^{2\Lambda}}{{\Phi'}}^{2}+{{e}^{2\Lambda}}{\Phi'}{\Lambda'}-4{{e}^{2\Lambda}}{{\varphi_0'}}^{2}+3{\varphi_0'}{\varphi_0''}r-{{e}^{2\Lambda}}{\Phi''}+{{\Phi'}}^{2}
\nonumber \\
&&-3{\Phi'}{\Lambda'}+3{{\varphi_0'}}^{2}+{\Phi''} 
\bigg] 
\left({{\frac{d^2f}{d\varphi_0^2}}}\right)
+2{r}^{4}{{e}^{6\Lambda}}{{\varphi_0'}}^{3} \left( {{e}^{2\Lambda}}-1 \right) \left({{\frac{d^3f}{d\varphi_0^3}}}\right)
\nonumber \\
&&+2{r}^{2}{{e}^{6\Lambda}} 
\left. \bigg[
10{{e}^{2\Lambda}}{{\varphi_0'}}^{3}{r}^{2}+4{\Lambda'}{{\varphi_0'}}^{3}{r}^{3}-4{\Phi'}{{\varphi_0'}}^{3}{r}^{3}-12{{\varphi_0'}}^{2}{\varphi_0''}{r}^{3}
+4{\Phi'}{\varphi_0''}{r}^{2}+3{\Phi'}{\varphi_0'}r
\right. 
\nonumber \\ 
&&\left. 
+2{\Lambda'}{\varphi_0'}r-30{{\varphi_0'}}^{3}{r}^{2}-{{e}^{4\Lambda}}{\varphi_0'} 
+2{{e}^{2\Lambda}}{\varphi_0'}-{\varphi_0''}r+{{e}^{2\Lambda}}{\varphi_0''}r+4{{\Phi'}}^{2}{\varphi_0'}{r}^{2}+3{\Phi''}{\varphi_0'}{r}^{2}
\right. 
\nonumber \\ 
&&\left. 
+2{{e}^{2\Lambda}}{\Phi'}{\Lambda'}{\varphi_0'}{r}^{2}-{\varphi_0'}-2{{e}^{2\Lambda}}{\Phi'}{\varphi_0''}{r}^{2}-2{{e}^{2\Lambda}}{\Phi'}{\varphi_0'}r-14{\Phi'}{\Lambda'}{\varphi_0'}{r}^{2}
+{{e}^{4\Lambda}}{\Phi'}{\varphi_0'}r
\right. 
\nonumber \\ 
&&\left. 
+2{{e}^{2\Lambda}}{\Phi'}{{\varphi_0'}}^{3}{r}^{3}-2{{e}^{2\Lambda}}{\Lambda'}{{\varphi_0'}}^{3}{r}^{3}+2{{e}^{2\Lambda}}{{\varphi_0'}}^{2}{\varphi_0''}{r}^{3}-2{{e}^{2\Lambda}}{{\Phi'}}^{2}{\varphi_0'}{r}^{2}-{{e}^{2\Lambda}}{\Phi''}{\varphi_0'}{r}^{2} \right. \bigg] \left({{\frac{df}{d\varphi_0}}}\right),
\end{eqnarray}
\begin{eqnarray}
D_4&=&-4{{\varphi_0'}}^{4}{r}^{3} \left( {{e}^{8\Lambda}}-4{{e}^{6\Lambda}}+3{{e}^{4\Lambda}} \right) {\left({{\frac{d^2f}{d\varphi_0^2}}}\right)}^{2} \nonumber \\ 
&&+ \bigg[ 
-8{{e}^{8\Lambda}}{\Phi'}{\Lambda'}{\varphi_0'}{r}^{2}-16{{e}^{6\Lambda}}{\Phi'}{\Lambda'}{\varphi_0'}{r}^{2}+8{{e}^{4\Lambda}}{\Phi'}{\Lambda'}{\varphi_0'}{r}^{2}-8{\varphi_0''}{{e}^{6\Lambda}}r+4{{e}^{8\Lambda}}{\varphi_0''}r
\nonumber \\
&&-24{{e}^{4\Lambda}}{{\varphi_0'}}^{5}{r}^{4}-4{{e}^{4\Lambda}}{\varphi_0'}+4{{e}^{8\Lambda}}{\varphi_0'}+4{\varphi_0'}{{e}^{6\Lambda}}-4{\varphi_0'}{{e}^{10\Lambda}}+4{{e}^{4\Lambda}}{\varphi_0''}r+36{{e}^{8\Lambda}}{{\varphi_0'}}^{3}{r}^{2}
\nonumber \\
&&-104{{e}^{6\Lambda}}{{\varphi_0'}}^{3}{r}^{2}+68{{e}^{4\Lambda}}{{\varphi_0'}}^{3}{r}^{2}+8{{e}^{6\Lambda}}{{\varphi_0'}}^{5}{r}^{4}-56{{e}^{4\Lambda}}{\Lambda'}{{\varphi_0'}}^{3}{r}^{3}+8{{e}^{8\Lambda}}{{\varphi_0'}}^{2}{\varphi_0''}{r}^{3}
\nonumber \\
&&-56{{e}^{6\Lambda}}{{\varphi_0'}}^{2}{\varphi_0''}{r}^{3}+48{{e}^{4\Lambda}}{{\varphi_0'}}^{2}{\varphi_0''}{r}^{3}+8{{e}^{8\Lambda}}{{\Phi'}}^{2}{\varphi_0'}{r}^{2}-32{{e}^{6\Lambda}}{{\Phi'}}^{2}{\varphi_0'}{r}^{2}+24{{e}^{4\Lambda}}{{\Phi'}}^{2}{\varphi_0'}{r}^{2}
\nonumber \\
&&+8{{e}^{8\Lambda}}{\Phi''}{\varphi_0'}{r}^{2}-24{{e}^{6\Lambda}}{\Phi''}{\varphi_0'}{r}^{2}
+16{{e}^{6\Lambda}}{\Phi'}{\varphi_0''}{r}^{2}+16{{e}^{4\Lambda}}{\Phi''}{\varphi_0'}{r}^{2}-16{{e}^{4\Lambda}}{\Phi'}{\varphi_0''}{r}^{2}+8{{e}^{8\Lambda}}{\Phi'}{\varphi_0'}r
\nonumber \\
&&-16{{e}^{4\Lambda}}{\Lambda'}{\varphi_0'}r
-8{{e}^{8\Lambda}}{\Lambda'}{{\varphi_0'}}^{3}{r}^{3}-88{{e}^{6\Lambda}}{\Phi'}{{\varphi_0'}}^{3}{r}^{3}+32{{e}^{6\Lambda}}{\Lambda'}{{\varphi_0'}}^{3}{r}^{3}+124{{e}^{4\Lambda}}{\Phi'}{{\varphi_0'}}^{3}{r}^{3}
\nonumber \\
&&-32{\Phi'}{\varphi_0'}{{e}^{6\Lambda}}r+16{\Lambda'}{\varphi_0'}{{e}^{6\Lambda}}r+12{{e}^{8\Lambda}}{\Phi'}{{\varphi_0'}}^{3}{r}^{3}+24{{e}^{4\Lambda}}{\Phi'}{\varphi_0'}r \bigg] \left({{\frac{df}{d\varphi_0}}}\right)
\left({{\frac{d^2f}{d\varphi_0^2}}}\right)
\nonumber \\
&&+4{{\varphi_0'}}^{2}r \bigg[ -6{{\varphi_0'}}^{2}{{e}^{6\Lambda}}{r}^{2}+{{e}^{8\Lambda}}{{\varphi_0'}}^{2}{r}^{2}+5{{e}^{4\Lambda}}{{\varphi_0'}}^{2}{r}^{2}+4{\Phi'}{{e}^{6\Lambda}}r-4{{e}^{4\Lambda}}{\Phi'}r-2{{e}^{6\Lambda}}+{{e}^{8\Lambda}} 
\nonumber \\ 
&&+{{e}^{4\Lambda}}
 \bigg] \left({{\frac{df}{d\varphi_0}}}\right)\left({{\frac{d^3f}{d\varphi_0^3}}}\right)
+ \bigg[ 
8{{e}^{8\Lambda}}{\Phi'}{\Lambda'}{{\varphi_0'}}^{2}{r}^{3}+8{{e}^{6\Lambda}}{\Phi'}{\Lambda'}{{\varphi_0'}}^{2}{r}^{3}+80{{e}^{4\Lambda}}{\Phi'}{\Lambda'}{{\varphi_0'}}^{2}{r}^{3}
\nonumber \\
&&-4{{e}^{8\Lambda}}{\Phi'}{\varphi_0'}{\varphi_0''}{r}^{3}-16{{e}^{6\Lambda}}{\Phi'}{\varphi_0'}{\varphi_0''}{r}^{3}+4{{e}^{4\Lambda}}{\Phi'}{\varphi_0'}{\varphi_0''}{r}^{3}
+4{{e}^{4\Lambda}}{\Phi'}+24{{e}^{4\Lambda}}{{\varphi_0'}}^{2}r
\nonumber \\
&&+24{{e}^{6\Lambda}}{{\Phi'}}^{2}r+168{{e}^{4\Lambda}}{{\varphi_0'}}^{4}{r}^{3}+4{{e}^{10\Lambda}}{\Phi'}
-4{{e}^{8\Lambda}}{\Phi'}-4{{e}^{6\Lambda}}{\Phi'}+8{{e}^{8\Lambda}}{{\varphi_0'}}^{4}{r}^{3}
-32{{e}^{6\Lambda}}{{\varphi_0'}}^{2}r
\nonumber \\
&&-8{{e}^{8\Lambda}}{{\Phi'}}^{2}r
+8{{e}^{6\Lambda}}{\Phi''}r-4{{e}^{8\Lambda}}{\Phi''}r-8{{e}^{6\Lambda}}{{\Phi'}}^{3}{r}^{2}-16{{e}^{4\Lambda}}{{\Phi'}}^{2}r+8{{e}^{4\Lambda}}{{\Phi'}}^{3}{r}^{2}
\nonumber \\
&&-4{{e}^{4\Lambda}}{\Phi''}r-80{{e}^{6\Lambda}}{{\varphi_0'}}^{4}{r}^{3}+8{{e}^{8\Lambda}}{{\varphi_0'}}^{2}r-16{{e}^{6\Lambda}}{\Phi'}{{\varphi_0'}}^{4}{r}^{4}
+16{{e}^{6\Lambda}}{\Lambda'}{{\varphi_0'}}^{4}{r}^{4}-16{{e}^{6\Lambda}}{{\varphi_0'}}^{3}{\varphi_0''}{r}^{4}
\nonumber \\
&&+48{{e}^{4\Lambda}}{{\varphi_0'}}^{3}{\varphi_0''}{r}^{4}
-8{{e}^{8\Lambda}}{{\Phi'}}^{2}{{\varphi_0'}}^{2}{r}^{3}+16{{e}^{6\Lambda}}{{\Phi'}}^{2}{{\varphi_0'}}^{2}{r}^{3}-40{{e}^{4\Lambda}}{{\Phi'}}^{2}{{\varphi_0'}}^{2}{r}^{3}
-4{{e}^{8\Lambda}}{\Phi''}{{\varphi_0'}}^{2}{r}^{3}
\nonumber \\
&&+16{{e}^{6\Lambda}}{\Phi''}{{\varphi_0'}}^{2}{r}^{3}-44{{e}^{4\Lambda}}{\Phi''}{{\varphi_0'}}^{2}{r}^{3}
-24{{e}^{8\Lambda}}{\Phi'}{{\varphi_0'}}^{2}{r}^{2}-24{{e}^{8\Lambda}}{\Lambda'}{{\varphi_0'}}^{2}{r}^{2}+40{{e}^{6\Lambda}}{{\Phi'}}^{2}{\Lambda'}{r}^{2}
\nonumber \\
&&+104{{e}^{6\Lambda}}{\Phi'}{{\varphi_0'}}^{2}{r}^{2}+8{{e}^{6\Lambda}}{\Lambda'}{{\varphi_0'}}^{2}{r}^{2}-56{{e}^{4\Lambda}}{{\Phi'}}^{2}{\Lambda'}{r}^{2}
-128{{e}^{4\Lambda}}{\Phi'}{{\varphi_0'}}^{2}{r}^{2}+8{{e}^{8\Lambda}}{\varphi_0'}{\varphi_0''}{r}^{2}
\nonumber \\
&&-16{{e}^{6\Lambda}}{\Phi''}{\Phi'}{r}^{2}
-8{{e}^{6\Lambda}}{\varphi_0'}{\varphi_0''}{r}^{2}+16{{e}^{4\Lambda}}{\Phi''}{\Phi'}{r}^{2}-16{{e}^{6\Lambda}}{\Phi'}{\Lambda'}r+16{{e}^{4\Lambda}}{\Phi'}{\Lambda'}r \bigg]
 {\left({{\frac{df}{d\varphi_0}}}\right)}^{2},
\end{eqnarray}
\begin{eqnarray}
D_6&=&8{{e}^{2\Lambda}}{{\varphi_0'}}^{3} 
\bigg[
2{{e}^{4\Lambda}}{{\varphi_0'}}^{2}{r}^{2}-8{{e}^{2\Lambda}}{{\varphi_0'}}^{2}{r}^{2}+6{{\varphi_0'}}^{2}{r}^{2}-12{{e}^{4\Lambda}}{\Phi'}r+32{{e}^{2\Lambda}}{\Phi'}r-20{\Phi'}r
 \nonumber \\
&&+{{e}^{6\Lambda}}+{{e}^{4\Lambda}}-5{{e}^{2\Lambda}}+3 
\bigg]
 \left({{\frac{df}{d\varphi_0}}}\right){\left({{\frac{d^2f}{d\varphi_0^2}}}\right)}^{2}
+8{{e}^{2\Lambda}}{\varphi_0'} 
\bigg[ 
 -9{\varphi_0'}{\Phi'}+2{{e}^{2\Lambda}}{\varphi_0''}
\nonumber \\
 &&-4{{e}^{4\Lambda}}{\varphi_0''}+10{{e}^{4\Lambda}}{\Lambda'}{\varphi_0'}+{\varphi_0'}{\Phi'}{{e}^{6\Lambda}}
 -6{\Lambda'}{\varphi_0'}{{e}^{6\Lambda}}-{{e}^{6\Lambda}}{\Phi'}{\Lambda'}{\varphi_0'}r+13{{e}^{4\Lambda}}{\Phi'}{\Lambda'}{\varphi_0'}r
 \nonumber \\ 
 &&-11{{e}^{2\Lambda}}{\Phi'}{\Lambda'}{\varphi_0'}r
 +12{\Lambda'}{{\varphi_0'}}^{3}{r}^{2}-2{\Phi'}{\varphi_0''}r-18{{e}^{4\Lambda}}{{\varphi_0'}}^{3}r+2{{e}^{6\Lambda}}{{\varphi_0'}}^{3}r-30{{\varphi_0'}}^{3}r
 \nonumber \\ 
 &&+6{\Lambda'}{\varphi_0'}
 -4{{e}^{4\Lambda}}{{\varphi_0'}}^{2}{\varphi_0''}{r}^{2}+16{{e}^{2\Lambda}}{{\varphi_0'}}^{2}{\varphi_0''}{r}^{2}-10{{e}^{4\Lambda}}{\Phi'}{{\varphi_0'}}^{3}{r}^{2}
+4{{e}^{4\Lambda}}{\Lambda'}{{\varphi_0'}}^{3}{r}^{2} 
 \nonumber \\ 
 &&+24{{e}^{2\Lambda}}{\Phi'}{{\varphi_0'}}^{3}{r}^{2}-8{{e}^{2\Lambda}}{\Lambda'}{{\varphi_0'}}^{3}{r}^{2}
 +3{{e}^{4\Lambda}}{{\Phi'}}^{2}{\varphi_0'}r-9{{e}^{2\Lambda}}{{\Phi'}}^{2}{\varphi_0'}r+{{\Phi'}}^{2}{\varphi_0'}{{e}^{6\Lambda}}r
 \nonumber \\ 
&&-9{{e}^{4\Lambda}}{\Phi''}{\varphi_0'}r
 -10{{e}^{4\Lambda}}{\Phi'}{\varphi_0''}r+11{{e}^{2\Lambda}}{\Phi''}{\varphi_0'}r+12{{e}^{2\Lambda}}{\Phi'}{\varphi_0''}r+{\Phi''}{\varphi_0'}{{e}^{6\Lambda}}r
-9{\Phi'}{\Lambda'}{\varphi_0'}r
 \nonumber \\ 
&& +11{{e}^{2\Lambda}}{\Phi'}{\varphi_0'}-3{{e}^{4\Lambda}}{\Phi'}{\varphi_0'}+2{\varphi_0''}{{e}^{6\Lambda}}-3{\Phi''}{\varphi_0'}r
-12{{\varphi_0'}}^{2}{\varphi_0''}{r}^{2} 
-10{{e}^{2\Lambda}}{\Lambda'}{\varphi_0'}+46{{e}^{2\Lambda}}{{\varphi_0'}}^{3}r 
 \nonumber \\ 
&& -30{\Phi'}{{\varphi_0'}}^{3}{r}^{2}+5{{\Phi'}}^{2}{\varphi_0'}r 
\bigg]
  {\left({{\frac{df}{d\varphi_0}}}\right)}^{2}\left({{\frac{d^2f}{d\varphi_0^2}}}\right)
+8{{e}^{2\Lambda}}{{\varphi_0'}}^{3} 
\bigg[
   -2{{e}^{4\Lambda}}{{\varphi_0'}}^{2}{r}^{2}
+8{{e}^{2\Lambda}}{{\varphi_0'}}^{2}{r}^{2}
 \nonumber \\  
&&  -6{{\varphi_0'}}^{2}{r}^{2}+2{{e}^{4\Lambda}}{\Phi'}r-20{{e}^{2\Lambda}}{\Phi'}r
   +18{\Phi'}r+{{e}^{6\Lambda}}-5{{e}^{4\Lambda}}+7{{e}^{2\Lambda}}-3 
\bigg]
    {\left({{\frac{df}{d\varphi_0}}}\right)}^{2}\left({{\frac{d^3f}{d\varphi_0^3}}}\right)
     \nonumber \\
    &&-8{{e}^{2\Lambda}} 
\bigg[
     3{\Phi'}{\varphi_0''}+{{e}^{4\Lambda}}{\Phi'}{\varphi_0''}+7{{e}^{2\Lambda}}{\Phi''}{\varphi_0'}-2{{e}^{2\Lambda}}{{\Phi'}}^{2}{\varphi_0'}
-5{{e}^{2\Lambda}}{\Phi'}{\varphi_0''}+{\Phi'}{\varphi_0''}{{e}^{6\Lambda}}
 \nonumber \\     
&& +2{{\Phi'}}^{2}{\varphi_0'}{{e}^{6\Lambda}}
     -5{{e}^{4\Lambda}}{\Phi''}{\varphi_0'}  
     +6{\Phi'}{\Lambda'}{\varphi_0'}-16{{e}^{4\Lambda}}{\Phi'}{\Lambda'}{{\varphi_0'}}^{3}{r}^{2}+4{{e}^{2\Lambda}}{\Phi'}{\Lambda'}{{\varphi_0'}}^{3}{r}^{2}
 \nonumber \\     
&& +6{{e}^{4\Lambda}}{\Phi'}{{\varphi_0'}}^{2}{\varphi_0''}{r}^{2} 
    -12{{e}^{2\Lambda}}{\Phi'}{{\varphi_0'}}^{2}{\varphi_0''}{r}^{2}+10{{e}^{4\Lambda}}{{\Phi'}}^{2}{\Lambda'}{\varphi_0'}r
 -8{{e}^{2\Lambda}}{{\Phi'}}^{2}{\Lambda'}{\varphi_0'}r
 -4{{e}^{4\Lambda}}{\Phi''}{\Phi'}{\varphi_0'}r 
 \nonumber \\     
 &&  -8{{e}^{2\Lambda}}{\Phi''}{\Phi'}{\varphi_0'}r-6{\Phi''}{{\varphi_0'}}^{3}{r}^{2}-14{{\Phi'}}^{2}{\varphi_0''}r+6{{\Phi'}}^{3}{\varphi_0'}r-3{\Phi''}{\varphi_0'}
-10{{e}^{4\Lambda}}{\Phi'}{{\varphi_0'}}^{3}r
 \nonumber \\     
&&     -4{{e}^{2\Lambda}}{{\Phi'}}^{3}{\varphi_0'}r+22{{e}^{2\Lambda}}{\Phi'}{{\varphi_0'}}^{3}r-10{{e}^{4\Lambda}}{{\Phi'}}^{2}{\varphi_0''}r
+24{{e}^{2\Lambda}}{{\Phi'}}^{2}{\varphi_0''}r-8{{e}^{2\Lambda}}{{\Phi'}}^{2}{{\varphi_0'}}^{3}{r}^{2}
 \nonumber \\     
&&     +12{\Phi'}{\Lambda'}{{\varphi_0'}}^{3}{r}^{2}+6{\Phi'}{{\varphi_0'}}^{2}{\varphi_0''}{r}^{2}    
     -18{{\Phi'}}^{2}{\Lambda'}{\varphi_0'}r+12{\Phi''}{\Phi'}{\varphi_0'}r+2{{e}^{4\Lambda}}{\Phi''}{{\varphi_0'}}^{3}{r}^{2}-4{{e}^{2\Lambda}}{\Phi''}{{\varphi_0'}}^{3}{r}^{2}
 \nonumber \\     
&&     +2{{e}^{6\Lambda}}{\Phi'}{{\varphi_0'}}^{3}r-2{{e}^{4\Lambda}}{{\Phi'}}^{3}{\varphi_0'}r-6{\Phi'}{\Lambda'}{\varphi_0'}{{e}^{6\Lambda}}+10{{e}^{4\Lambda}}{\Phi'}{\Lambda'}{\varphi_0'}
 \nonumber \\     
&&     -10{{e}^{2\Lambda}}{\Phi'}{\Lambda'}{\varphi_0'}+{\Phi''}{\varphi_0'}{{e}^{6\Lambda}}-30{\Phi'}{{\varphi_0'}}^{3}r
\bigg]
      {\left({{\frac{df}{d\varphi_0}}}\right)}^{3},
\end{eqnarray}
\begin{eqnarray}
D_8&=&-32{{\varphi_0'}}^{4}{\Phi'} 
\bigg[ 
3{{e}^{6\Lambda}}-13{{e}^{4\Lambda}}+25{{e}^{2\Lambda}}-15 
\bigg]
{\left({{\frac{df}{d\varphi_0}}}\right)}^{2}{\left({{\frac{d^2f}{d\varphi_0^2}}}\right)}^{2}
\nonumber \\
&&-32{{\varphi_0'}}^{4}{\Phi'} \left( {{e}^{6\Lambda}}-{{e}^{4\Lambda}}-9{{e}^{2\Lambda}}+9 \right) {\left({{\frac{df}{d\varphi_0}}}\right)}^{3}\left({{\frac{d^3f}{d\varphi_0^3}}}\right)
-64{{\varphi_0'}}^{2} 
\bigg[
-{{\Phi'}}^{2}{\varphi_0'}{{e}^{6\Lambda}}
\nonumber \\
&&-5{\Phi'}{\Lambda'}{\varphi_0'}{{e}^{6\Lambda}}
+3{{e}^{4\Lambda}}{{\Phi'}}^{2}{\varphi_0'}+10{{e}^{4\Lambda}}{\Phi'}{\Lambda'}{\varphi_0'}
-5{{e}^{2\Lambda}}{{\Phi'}}^{2}{\varphi_0'}
-9{{e}^{2\Lambda}}{\Phi'}{\Lambda'}{\varphi_0'}
\nonumber \\
&&+{\Phi''}{\varphi_0'}{{e}^{6\Lambda}}
+2{\Phi'}{\varphi_0''}{{e}^{6\Lambda}}
-4{{e}^{4\Lambda}}{\Phi''}{\varphi_0'}-7{{e}^{4\Lambda}}{\Phi'}{\varphi_0''}+3{{e}^{2\Lambda}}{\Phi''}{\varphi_0'}
+8{{e}^{2\Lambda}}{\Phi'}{\varphi_0''}
\nonumber \\
&&+3{{\Phi'}}^{2}{\varphi_0'}
-3{\Phi'}{\varphi_0''} 
\bigg]
 {\left({{\frac{df}{d\varphi_0}}}\right)}^{3}\left({{\frac{d^2f}{d\varphi_0^2}}}\right)
+64{\varphi_0'}{\Phi'} 
\bigg[
{{\Phi'}}^{2}{\varphi_0'}{{e}^{6\Lambda}}
-5{\Phi'}{\Lambda'}{\varphi_0'}{{e}^{6\Lambda}}
\nonumber \\
&&-4{{e}^{4\Lambda}}{{\Phi'}}^{2}{\varphi_0'}+10{{e}^{4\Lambda}}{\Phi'}{\Lambda'}{\varphi_0'}
+3{{e}^{2\Lambda}}{{\Phi'}}^{2}{\varphi_0'}
-9{{e}^{2\Lambda}}{\Phi'}{\Lambda'}{\varphi_0'}+2{\Phi''}{\varphi_0'}{{e}^{6\Lambda}}
\nonumber \\
&&+{\Phi'}{\varphi_0''}{{e}^{6\Lambda}}
-8{{e}^{4\Lambda}}{\Phi''}{\varphi_0'}-3{{e}^{4\Lambda}}{\Phi'}{\varphi_0''}+6{{e}^{2\Lambda}}{\Phi''}{\varphi_0'}+5{{e}^{2\Lambda}}{\Phi'}{\varphi_0''}-3{\Phi'}{\varphi_0''} 
\bigg]
 {\left({{\frac{df}{d\varphi_0}}}\right)}^{4} .
\end{eqnarray}

\bibliographystyle{unsrt}

\end{document}